\documentclass[acmsmall]{acmart}
\acmJournal{PACMHCI}

\AtBeginDocument{%
  \providecommand\BibTeX{{%
    \normalfont B\kern-0.5em{\scshape i\kern-0.25em b}\kern-0.8em\TeX}}}
\usepackage{algorithm}
\usepackage{algpseudocode}
\usepackage{float}
\usepackage{subfigure}
\AtBeginDocument{%
  \providecommand\BibTeX{{%
    \normalfont B\kern-0.5em{\scshape i\kern-0.25em b}\kern-0.8em\TeX}}}
\usepackage{tabularx}
\usepackage{bbding}
\usepackage{tcolorbox}
\usepackage{fancyvrb}
\usepackage{graphicx}
\usepackage{geometry}
\usepackage{amsmath}
\usepackage{makecell}
\usepackage{float}
\usepackage{color}
\usepackage{booktabs}
\usepackage{caption}
\usepackage{tabu}
\usepackage{hyperref}
\usepackage{todonotes}
\usepackage{booktabs}
\usepackage{multirow}
\usepackage{graphicx}
\usepackage{subcaption}
\usepackage{soul}
\usepackage{booktabs,longtable}
\usepackage{booktabs} % For professional looking tables
\usepackage{tabularx} % For adjustable column widths
\usepackage{ragged2e} % For better text alignment within cells
\usepackage[utf8]{inputenc}
\usepackage[T1]{fontenc}
\usepackage{textgreek}

\usepackage{graphicx}
\usepackage{tabularray}

\usepackage{adjustbox} 

\definecolor{lightred}{HTML}{D89090}
\definecolor{darkblue}{HTML}{104E8B}
\definecolor{midblue}{HTML}{376B9E}
\definecolor{lightblue}{HTML}{5F89B1}
\definecolor{newgreen}{HTML}{C5E9E3}

\usepackage[dvipsnames]{xcolor} % keep xcolor, no need for plain color separately
\usepackage[normalem]{ulem}     % for \sout (strikeout)

\newif\iffinalversion
\finalversiontrue         % <<< FINAL version (clean, black text)

\iffinalversion
  \newcommand{\deleted}[1]{}       % remove deleted text completely
\else
  \newcommand{\deleted}[1]{\textcolor{red}{\sout{#1}}}
\fi

\AtBeginDocument{%
  \providecommand\BibTeX{{%
    Bib\TeX}}}

\begin{document}

\author{Jingfei Huang}
\authornote{These authors contributed equally to this work.}
\orcid{0009-0002-0213-4160}
\affiliation{
  \institution{Northeastern University}
    % \department{Graduate School of Design \& School of Engineering and Applied Science}
  \city{Boston}
  \state{Massachusetts}
  \country{USA}
}
\email{huang.jingfe@northeastern.edu}

\author{Yuyao Zhang}
\authornotemark[1]
\orcid{0009-0002-1531-3408}
\affiliation{
  \institution{The Hong Kong University of Science and Technology}
 % \department{School of Creative Media}
  \city{Hong Kong, SAR}
  \country{China}
}
\email{yzhang075@connect.ust.hk}

\author{Ruyan Chen}
\authornotemark[1]
\orcid{0009-0008-7660-8043}
\affiliation{
  \institution{City University of Hong Kong Studio for Narrative Spaces}
  \city{Hong Kong, SAR}
  \country{China}
}
\email{rc4514@nyu.edu}

%do not change.
\author{RAY LC}
\authornote{Correspondence can be addressed to ray.lc@cityu.edu.hk.}
\email{ray.lc@cityu.edu.hk}
\orcid{0000-0001-7310-8790}
\affiliation{
\institution{City University of Hong Kong Studio for Narrative Spaces}
\city{Hong Kong, SAR}
\country{China}}

\renewcommand{\shortauthors}{Huang, et al.}

%% Rights management information.  This information is sent to you
%% when you complete the rights form.  These commands have SAMPLE
%% values in them; it is your responsibility as an author to replace
%% the commands and values with those provided to you when you
%% complete the rights form.
%% Rights management information.  This information is sent to you
%% when you complete the rights form. CHANGE THIS:
\setcopyright{cc}
\setcctype{by-nc-nd}
\acmJournal{PACMHCI}
\acmYear{2026} \acmVolume{10} \acmNumber{7} \acmArticle{GAMES036}
\acmMonth{11} \acmDOI{10.1145/3831296}

%%\citestyle{authoryear}
\title[Hakka Kitchen]{Hakka Kitchen: Engagement with Culinary Cultural Heritage Through Immersive Game Play}

\begin{abstract}
%ver 2026-07-15 final edit rlc
Intangible Cultural Heritage (ICH) experiences are difficult to share with the public because they are essentially processes that rely on physical interactions with embodied, tacit, and situated phenomena in specific cultural contexts. We consume non-interactive media such as videos and books to learn about culinary ICH experiences, but they do not allow us to grasp actual interactive procedures that embody the cultural knowledge. To engage people in a traditional cooking experience, we created a gamified VR experience Hakka Kitchen, where players are guided by a chef of Hakka cuisine through a modeled physical process of making the traditional dish of stuffed bitter melon. Compared against watching a video in VR providing the same information in a between-subjects study (N=40), Hakka Kitchen led to increased sensory, imaginative engagement, positive affect, and willingness to transmit awareness for the culinary ICH. Heritage recognition is mode-dependent: the cognitive load of enactment suppressed players’ uptake of concurrent cultural narration; however, it generated somatic and memory anchors that fostered a deeper, implicit kinesthetic form of empathy for the artisan's labor and cultural background. Our work shows how representing interactive procedures instead of static content may empower cultural awareness in a virtual form of ICH practice.
\end{abstract}

% process - procedure to make specific things, and context - environment; 

%%
%% The code below is copied from, generated by the tool at http://dl.acm.org/ccs.cfm.
\begin{CCSXML}
<ccs2012>
   <concept>
       <concept_id>10003120.10003130.10011762</concept_id>
       <concept_desc>Human-centered computing~Empirical studies in HCI</concept_desc>
       <concept_significance>500</concept_significance>
       </concept>
 </ccs2012>
\end{CCSXML}
\ccsdesc[500]{Human-centered computing~Empirical studies in HCI}

%%
%% Keywords.
\keywords{Edutainment design, Virtual reality, HCI, Intangible Cultural Heritage}

\begin{teaserfigure}
  \includegraphics[width=\textwidth]{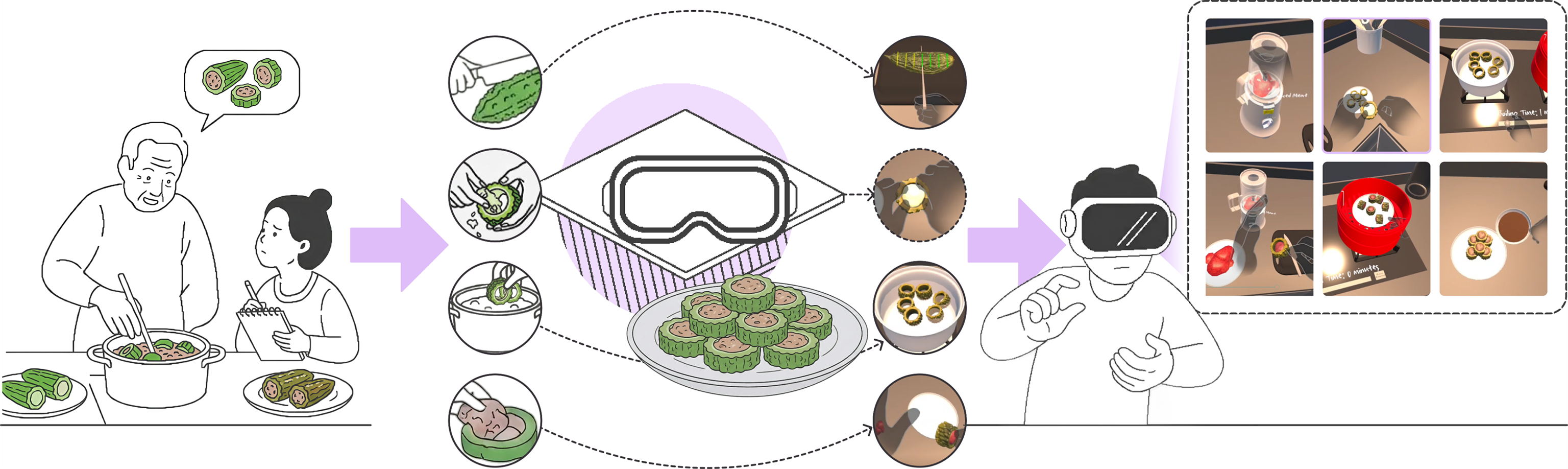}
  \caption{\textit{Hakka Kitchen} is a gamified VR experience that translates expert-interview insights into game mechanics, guiding players through preparing stuffed bitter melon under the mentorship of a Hakka chef.}
  \Description{.}
  \label{fig:teaser}
\end{teaserfigure}

% \centering
% \subfigure[]{
% \includegraphics[width=0.305\textwidth, trim= 0 0 40 3]{figs/cover1.JPG}\label{fig1a}
% }\hspace{1mm}
% \subfigure[]{
% \includegraphics[width=0.305\textwidth, trim= 0 0 50 0]{figs/cover2.JPG}\label{fig1b}
% }\hspace{1mm}
% \subfigure[]{
% \includegraphics[width=0.315\textwidth, trim= 0 0 50 0]{figs/cover3.JPG}
% \label{fig:01}
% }
% \caption{Caption}
% \Description{Caption}

\maketitle 

\section{Introduction}\label{sec:Introduction}
% Definition & Importance of ICH
Intangible Cultural Heritage (ICH) encompasses traditions or living expressions inherited from previous generations and transmitted onward~\cite{heritage2003}, constituting cultural identity and communal memory over time~\cite{zhao2022, zhang2024}. 
However, especially in the process-based culinary domain, ICH is difficult to safeguard at scale because its core lies in the embodied process of hands-on practice, knowledge that is physically understood yet resists verbal articulation and must be enacted to be transmitted~\cite{polanyi_tacit_2009, partarakis_representation_2021, rahmawati_acquisition_2015, farmer_tacit_2025}. 
Culinary heritage is also distinctive among embodied ICH: unlike traditions such as dance, kite-flying, or dragon-boat racing, whose meaning lies in expressive movement~\cite{heritage2003, sep-dance}, its cultural meaning is outcome-bearing and threshold-bound~\cite{bertinetto_dishes_2020, miton_cultural_2022} --- a dish either holds together or fails, and significance is encoded in narrow tolerances such as filling tightness.
% Limitations of Current Documentation Methods
Mainstream transmission nonetheless relies on documentary films, short videos, and recipe books~\cite{sun2024, zheng2023, bonn2016, wang_critical_2024}.
While valuable as information carriers, 
these methods position learners as observers rather than participants: passive viewing can transmit declarative knowledge but cannot exercise the sensorimotor contingencies 
through which cooking skill is actually formed~\cite{sutton2018, short2006, fjaeldstad2022, janhonen2018, feldman2023, rajan2023}.
Even when heritage content is accessed through interactive digital interfaces, interaction is often symbolically ``flattened'' to screen actions such as tapping, decoupling bodily gesture from culturally meaningful effects~\cite{ishii_tangible_1997, klemmer_how_2006}.
Thus, learners may gain familiarity with a tradition without experiencing the constraints through which its meanings are enacted.

% Potential & Shortcomings of Digital Games
Digital games have the potential to simulate complex procedural systems and provide structured feedback loops~\cite{kosmadoudi2013, garcia2019, buhalis2023, damavsevivcius2023}. 
While Augmented Reality (AR) and Leap Motion often face limitations in object manipulation or require physical props,
Virtual Reality (VR) supports such enactment through tracked, spatial interaction, enabling a more isomorphic mapping~\cite{lamas_head_2019, jacob_reality-based_2008}, a one-to-one spatial relationship between physical movements and virtual results~\cite{macaranas_what_2015}, than desktop or mobile inputs. 
This enables a digital ``apprenticeship'' in which users can physically enact the tacit dimensions of heritage~\cite{watanuki_knowledge_2007}.
Prior embodied VR and AR systems for ICH have already shown that such interaction raises engagement and cultural resonance relative to passive video, across domains including porcelain craft, folk dance, and traditional ceremony~\cite{ji_constructing_2021, su_embodied_2024, tan_case_2020, wu_kaibili_2025}.
However, what embodiment transmits, and at what cost, when heritage is constraint-bound rather than expressive remains an open question. 
Existing VR culinary applications typically prioritize entertainment~\cite {gyaurov2022, schaffer2019}, motor rehabilitation~\cite{gorsic_cooperative_2018, pau_immersive_2023}, or efficiency~\cite{winanti_culinary_2024, bouck_playing_2025}, which optimize for game flow while effectively removing some difficulty that defines the artisan's skill.
By contrast, a gamification framing, by contrast, can make these constraints legible and sustain engagement with them~\cite{bellotti_serious_2012, habgood_motivating_2011, metcalfe2017}.

% Our study 
To address this, we introduce \textbf{\textit{Hakka Kitchen}}, a gamified VR experience that reconstructs the Hakka culinary ICH of ``stuffed bitter melon'' as a specific case study. 
Following \citet{Deterding_Dixon_Khaled_Nacke_2011}, we characterize it as a gamified VR experience: culturally grounded cooking practice scaffolded by game elements, in which physics-based hand interaction requires the player to perform the defining cultural gestures and satisfy their thresholds to progress.
Its somatic contribution is therefore not the repetition of steps, but the isomorphic enactment of threshold-bound constraints through the hand: for example, feeling with gestures and physics what a 1--3\,cm slice means, which symbolic media cannot convey.
We position \textit{Hakka Kitchen} not as a replacement for the material reality of a traditional kitchen where tactile feedback and authentic tools are important, but as a scalable tool for cultural popularization and awareness, to bridge tacit culinary knowledge with physical interaction, grounded in formative expert studies of the procedure and cultural narration, beyond the constraints of the scarcity of master chefs and safety risks.
Specifically, we investigate the following research questions:

\textbf{RQ1:} How can we translate the tacit dimensions of culinary ICH into explicit mechanics of a gamified VR experience?

 % (kinesthetic empathy)
\textbf{RQ2:} 
How do players interact with elements in the gamified experience when engaging with culinary ICH content?

\textbf{RQ3:} How do enactment and passive observation of culinary procedures in VR differ in which layers of culinary heritage they transmit when content is held constant?

% contribution - augment cultural awareness through game
Our contributions are:
(1) a generalizable process for documenting and translating tacit culinary heritage into embodied game mechanics. The process moves from semi-structured expert elicitation, to a chef-derived procedural dictionary that records each step's tacit checkpoint, somatic constraint, and cultural connotation, to its operationalization as interaction, gestures, and narration that externalize process-based know-how in situ;
We used stuffed bitter melon as the instance through which we demonstrate the process;
(2) a design framework that uses the ``enactment of constraints'' for culinary ICH practice, specifying how culturally grounded narration with step execution and using recoverable mistakes can preserve culturally important frictions in actions while remaining learnable and approachable;
(3) an empirical comparison between a gamified VR experience against a content-matched VR video, showing that interactive enactment raises  Sensory \& Imaginative Immersion, Positive Affect, and several dimensions of Heritage Awareness, and that recoverable mistakes shape engagement and perceived transfer while cultural narration reliably lands during observation but is partially masked by manual cognitive load during enactment, yielding design implications for future ICH transmission.

\section{Background}\label{sec:Background}
\subsection{Intangible Cultural Heritage} 
% The Challenge of "Tacit" Culinary Heritage
ICH encompasses the practices, representations, and expressions that communities recognize as their cultural legacy~\cite{heritage2003}.
Unlike tangible artifacts, ICH is enacted through ongoing human performance and social transmission~\cite{hou_digitizing_2022, partarakis_representation_2021}.  
Culinary ICH is particularly difficult to safeguard because its core lies in tacit knowledge, which is the embodied know-how that is physically understood but difficult to articulate verbally~\cite{polanyi_tacit_2009, partarakis_representation_2021, rahmawati_acquisition_2015, farmer_tacit_2025, lu2022}, therefore creating a transmission challenge~\cite{taylor_archive_2003}
that extends beyond recipe content to include the sensorimotor contingencies of the practitioner~\cite{oregan_sensorimotor_2001}.

% 2. The Failure of Current Media
% Case and Gap in Culinary ICH Preservation
% The Semantic Gap of Symbolic Interfaces
Current digital interventions prioritize observational media such as scans, photos, and 
videos that improve access but neglect procedural and affective layers~\cite{zhanna_digitization_2020, podara_digital_2021}. 
Large-scale platforms like \textit{Recipe1M+}~\cite{marin_recipe1m_2021} or \textit{FoodKG}~\cite{ghidini_foodkg_2019} and
social video platforms such as TikTok~\cite{paquienseguy_douyin_2025, wang_critical_2024} distribute heritage content for narrative absorption~\cite{hakemulder_narrative_2017},
% Gap
but they rely on symbolic input, such as tapping, clicking, that abstracts and flattens 
the physical gesture~\cite{ishii_tangible_1997, klemmer_how_2006}. 
The \textit{sensorimotor contingencies} a craft requires can only develop through imitative or enactive practice~\cite{chi_icap_2014, oregan_sensorimotor_2001, mian_comparing_2024}.
To bridge this gap, an interface must support \textit{isomorphic mapping}~\cite{macaranas_what_2015}, in which the user's physical input structurally resembles the cultural action, thereby triggering proprioceptive systems linked to muscle memory through hands-on practice~\cite{schwichow2016, nimkulrat2012}.

% serious game
Game-based learning embeds practice in situated contexts, often using narratives and generative systems~\cite{lc_chikyuchi_2022, zhou_eternagram_2024, zhang_can_2025} to influence awareness for social purpose~\cite{agcal_bricolage_2025, agcal_land_2023}. Immersive VR has been shown to raise motivation, perceived authenticity, and engagement for heritage content~\cite{qiu_continuance_2024, sun_application_2023, liu_evaluating_2025}.
% [NEW] other ICH in VR
Of these, Embodied Dance in Folk Cultural Space~\cite{su_embodied_2024} has non-expert learners embody folk dance in VR to enhance cultural experience. 
In this kind of movement-based heritage, however, cultural meaning is carried by the expressive form of the body in motion, and consequently, embodiment and cultural transmission are aligned: moving more is learning more.
In culinary ICH, the cultural narrative is a separate stream layered onto a manual task whose embodiment can crowd it out. 
Embodiment is therefore not uniformly additive for cultural uptake, creating a tension that movement-centric ICH does not surface. 
Extending prior VR-versus-video comparisons~\cite{ji_constructing_2021, nie_kites_2024}, we position our study as a characterization of what embodiment transmits, and at what cost, when the heritage is constraint-bound.
Other embodied VR and AR studies instantiate the same expression-bearing premise across heritage domains: 
embodied WebAR lets learners rehearse the maker's gestures to internalize techniques that video cannot convey~\cite{tan_case_2020};
gesture-based VR ceremony \textit{KaiBiLi}~\cite{wu_kaibili_2025} and multimodal embodied design for Cantonese dragon-boat culture~\cite{lei_study_2025} treat the body's movement as the primary carrier of meaning. 
We share their reliance on isomorphic input, but diverge in what that input must be faithful to.
Of this body of work, \textit{FloraJing}~\cite{wang_facilitating_2025}, a VR system for the daily practice of Traditional Chinese Flower Arrangement, which is a craft-based ICH, found that VR promoted progressive reflection and sustained cultural understanding over time. 
However, \textit{FloraJing} targeted existing practitioners who already possess the cultural frame and need scaffolded practice. 
This leaves a gap for culinary ICH practice designs accessible to participants across expertise levels.

% Consumer App
Consumer VR with hand-tracking provides a sufficiently high-fidelity environment to simulate craftsmanship~\cite{mohamad_cooking_2024}, 
to rehearse culturally relevant motor patterns and to enforce the constraint-driven feedback that observational media cannot.
\citet{habgood_motivating_2011} establish the governing principle that game-based learning succeeds when the learning goal is intrinsically integrated into the core mechanic, not bolted on as an external reward.
However, existing culinary video games such as \textit{Overcooked}~\cite{ghost_town_games_ltd_overcooked_2016}, \textit{Cooking Mama}~\cite{office_create_cooking_2020}, and \textit{Lost Recipes}~\cite{lostrecipes2022} prioritize usability and flow~\cite{sweetser_gameflow_2005, gorsic_cooperative_2018, pau_immersive_2023}, simplifying procedures and removing the real-world constraints where cultural meaning resides~\cite{lin_review_2025, mohamad_cooking_2024}. Then the constraint becomes an obstacle to be smoothed away rather than the thing being learned.
% Picking Stuffed Bitter Melon as a specific case
In \textit{Hakka Kitchen}, we address this by making those heritage constraints the mechanic itself, shifting the design goal from gamified efficiency to embodied cultural awareness and using a VR gamified experience to externalize the tacit constraints of the Hakka culinary tradition.

\subsection{Embodied Cognition}
\label{embodied_cognition}
% 1. Theory: From "Thinking" to "Enacting" (Enactivism)
% Effective Transmission Relies on Embodied Experience
Embodied cognition argues that cognition emerges from ongoing sensorimotor interaction between the body and its environment~\cite{shapiro2019embodied, foglia2013embodied, wilson2013embodied, de_jaegher_participatory_2007}.
Situated learning theories also argue that newcomers develop understanding through legitimate peripheral participation, acquiring the procedures, norms, attentional habits, and interpretive frames of the practice~\cite{lave_situated_1991, rogoff_apprenticeship_1990}. 
Heritage education further emphasizes that learning is contextual and affective: experiences that couple action, setting, and narrative support meaning-making and personal connection rather than simple information recall~\cite{falk_using_2005}.
In culinary practice, core knowledge is often enacted as perception-action loops, making bodily interaction central to how tacit know-how is learned and retained~\cite{koerich2024learning, askren2021, taylor_archive_2003}. 

% 2. The Mechanism: Kinesthetic Empathy (The "Why")
This enactment facilitates \textit{kinesthetic empathy}, the ability to understand another's experience by simulating their bodily actions~\cite{reynolds_kinesthetic_2012}, which is key to appreciating the cultural logic embedded in a practice.
For culinary ICH, the goal of immersive learning is not merely to improve procedural correctness but to create conditions in which learners can feel and interpret culturally meaningful constraints through guided enactment.

% 3. VR's Role: isomorphic mapping (The Solution)
VR supports this through \textit{isomorphic mapping}~\cite{macaranas_what_2015}.
% Limitations of Traditional Methods (Disembodiment)
Unlike current disembodied, passive methods for documenting and transmitting culinary ICH~\cite{plantinga2005, currie1999}, VR requires users to perform the somatic work of heritage.
% gap
% VR addresses this gap by enabling embodied simulation. 
Through head- and hand-tracking, VR immerses users in simulated culinary environments, thereby fostering the sense of spatial presence. 
Users can virtually manipulate tools and ingredients while linking actions to visual or auditory cues, which provides critical kinesthetic feedback. 
VR creates dynamic sensorimotor loops: actions (e.g., stirring) elicit sensory consequences (visual, sound, vibration changes), thereby requiring users to respond to approximate real-world feedback mechanisms for skill acquisition. 
Recent VR works such as \textit{Digital Diabolo}~\cite{Kong_2024} and \textit{ShadowPlayVR}~\cite{He_2023} demonstrate VR’s capacity to externalize tacit cultural techniques through embodied interaction.
These affordances motivate immersive designs for culinary ICH that aim not only at procedural correctness but also at cultural sense-making through embodied engagement; based on these, we designed \textit{Hakka Kitchen} to leverage embodied cognition to enhance experiential engagement and foster cultural awareness. 

Yet embodiment is not cost-free for every learning channel: 
performing a demanding manual task competes for limited cognitive resources with the concurrent encoding of other information~\cite{Chandler_Sweller_1991, Mayer_2022}, in particular, sustained sensorimotor coordination can suppress uptake of a simultaneous secondary stream~\cite{daniel_j_simons_gorillas_1999, tomporowski_cognitive-motor_2020}.
Multiple-resource accounts~\cite{wickens_multiple_2008} hold that this competition can be amodal. 
\textit{Action masking}, as we use the term, refers to the phenomenon whereby the motor demands of performing a procedural task suppress the learner's concurrent absorption of cultural narration: the body's engagement with the procedure monopolizes the attentional bandwidth needed to encode it~\cite{tomporowski_cognitive-motor_2020, daniel_j_simons_gorillas_1999, easterbrook_effect_1959}.
This sets up a tension specific to constraint-bound culinary ICH, where the somatic layer (felt constraints) and the symbolic layer (cultural narrative) must be acquired together, which is a possibility we examine empirically, and one that movement-centric ICH does not confront.

\subsection{Culinary ICH Design}
Within VR heritage interaction, the dominant approach centers on visually driven 3D reconstruction of sites and artifacts, with experiences designed as virtual tours or curatorial object inspection~\cite{innocente_framework_2023, rodriguez-garcia_systematic_2024}.
Interaction in these systems is typically either exploratory or curatorial~\cite{tian_empirical_2024,komianos_immersive_2022}, prioritizing materiality over the intangible aspects of culture.
% GenAI
Recent work has addressed this by shifting the focus to narrative and co-creation, such as in immersive narrative spaces~\cite{miller_eliciting_2025} and in helping users recall the past by co-creating heritage artifacts~\cite{he_i_2025}, AI-VR for creative engagement~\cite{nie_kites_2024}, GenAI-driven narrative experiences~\cite{fu_being_2024, li_reality_2026}, and spatialized-audio interaction for performing-arts ICH~\cite{wang_temporal_2025}. These systems bridge the semantic gap at the symbolic level but not the somatic dimension.

Culinary practice is a well-established domain in HCI and XR, but it is typically treated as a proxy for general skill or coordination rather than as cultural heritage, with applications focused on tutoring and guidance~\cite{chen_holocook_2025, horie_interactive_2006, mega_assist_2008},
motor acquisition and rehabilitation~\cite{gorsic_cooperative_2018, pau_immersive_2023, lee2021}, 
cognitive therapy or training~\cite{zedda2021, kosch2019, liao_vr-cook_2026}, 
or entertainment flow~\cite{nakamoto2008, janter2023, gyaurov2022, schaffer2019}. 
% GEN AI
Recent work has engaged culinary ICH directly, but from adjacent angles, against which we position our work.
First, AI-mediated culinary-heritage work has aimed at creative adaptation and documentation. 
\textit{Salt is the Soul of Hakka Baked Chicken}~\cite{liu_salt_2025} found that GenAI co-design workshops with Hakka practitioners sparked recipe innovation but strained workflow accuracy and flavor fidelity, illustrating the creative-versus-fidelity tension in culinary ICH. 
\textit{Replacing Bak Choy with Cabbage}~\cite{zeng_replacing_2025} used GenAI as a participatory tool to help knowledgeable ICH-practitioners visualize alternative culinary scenarios and produce speculative recipe documentation. 
Second, prior works mainly focused on calibrated procedural engagement; for example, a GenAI-adaptive VR system for Neapolitan pizza-making showed that calibration improves procedural engagement without overwhelming learners~\cite{lau_adaptive_2025}, while food–technology research highlights the broader tension between digital efficiency and the artisanal, culturally situated nature of traditional cooking~\cite{balkaya_exploring_2025}. 
Both angles optimize for either creative variation or procedural fluency; however, whether embodied enactment of ICH constraints changes a learner's cultural understanding remains empirically unaddressed.

% why culinary ICH is a problem
Culinary ICH presents a distinct design problem, which is the tension between usability-driven friction removal and heritage-driven constraint preservation~\cite{metcalfe_memory_1994, kapur_productive_2008, sengers_reflective_2005, hallnas_slow_2001}, separating it from the movement-centric ICH in prior embodied-VR work, for three reasons. 
First, culinary ICH is outcome-bearing, and its cultural meaning is realized only when both the making process and the product succeed~\cite{unesco_food_2025, partarakis_representation_2021}; for example, a stuffed bitter melon that cannot hold its filling fails to embody the practice it represents.
Second, culinary ICH is threshold-bound, so its cultural meaning is also encoded in narrow tolerance bands, such as the 1--3\,cm slicing thickness, rather than in continuous expressive motion~\cite{farmer_tacit_2025, bohm_sensory_2026, sennett_craftsman_2008}. 
Third, it is chemosensory, such that the practitioner's judgments of doneness, texture, and flavor are grounded in sensory perception~\cite{pieniak_what_2023, bohm_sensory_2026, persky_olfactory_2020}. 
%
% Why not MultiSensory
Prior work in multisensory VR confirms that olfactory and gustatory cues remain the least-realized modalities in VR research~\cite{zholzhanova_virtual_2025}.  
Research-level systems show that chemosensory feedback is feasible:
\textit{ThermalGrasp}~\cite{mazursky_thermalgrasp_2024} delivers thermal sensation, 
% the buffet
a virtual buffet study demonstrates olfactory simulation in VR~\cite{persky_olfactory_2020},
and \textit{Vocktail}~\cite{ranasinghe_vocktail_2017} uses electrical and thermal tongue stimulation with smell and color to actuate taste. 
Specifically, \textit{VirCHEW Reality}~\cite{liu_virchew_2025} uses a face-worn pneumatic device to reproduce food texture without physical food in the mouth.
Including an additional sensory system in VR might better simulate realism; however, this requires specialized hardware incompatible with scalable deployment. 
Moreover, these systems simulate the consumption of a finished product, whereas culinary ICH meaning is carried in the constraint-laden making that precedes consumption. 
\textit{Hakka Kitchen} accepts this boundary condition and posits that isomorphic gesture-to-constraint mapping is sufficient for popularization and cultural awareness, while explicitly insufficient for vocational-fidelity transfer, framing our design goals and evaluation scope.
% Gap
This points to a design requirement for a culinary ICH gamified VR experience, such as \textit{Hakka Kitchen}: to intrinsically integrate the cultural learning goal with the core mechanic~\cite{habgood_motivating_2011},  treating embodied interaction as a medium for cultural sense-making: users experience culturally significant constraints as the game mechanics themselves.

\section{Game Design}\label{sec:Design}
\subsection{Overview}

\begin{figure}[htbp]
    \centering
    \includegraphics[width=\linewidth]{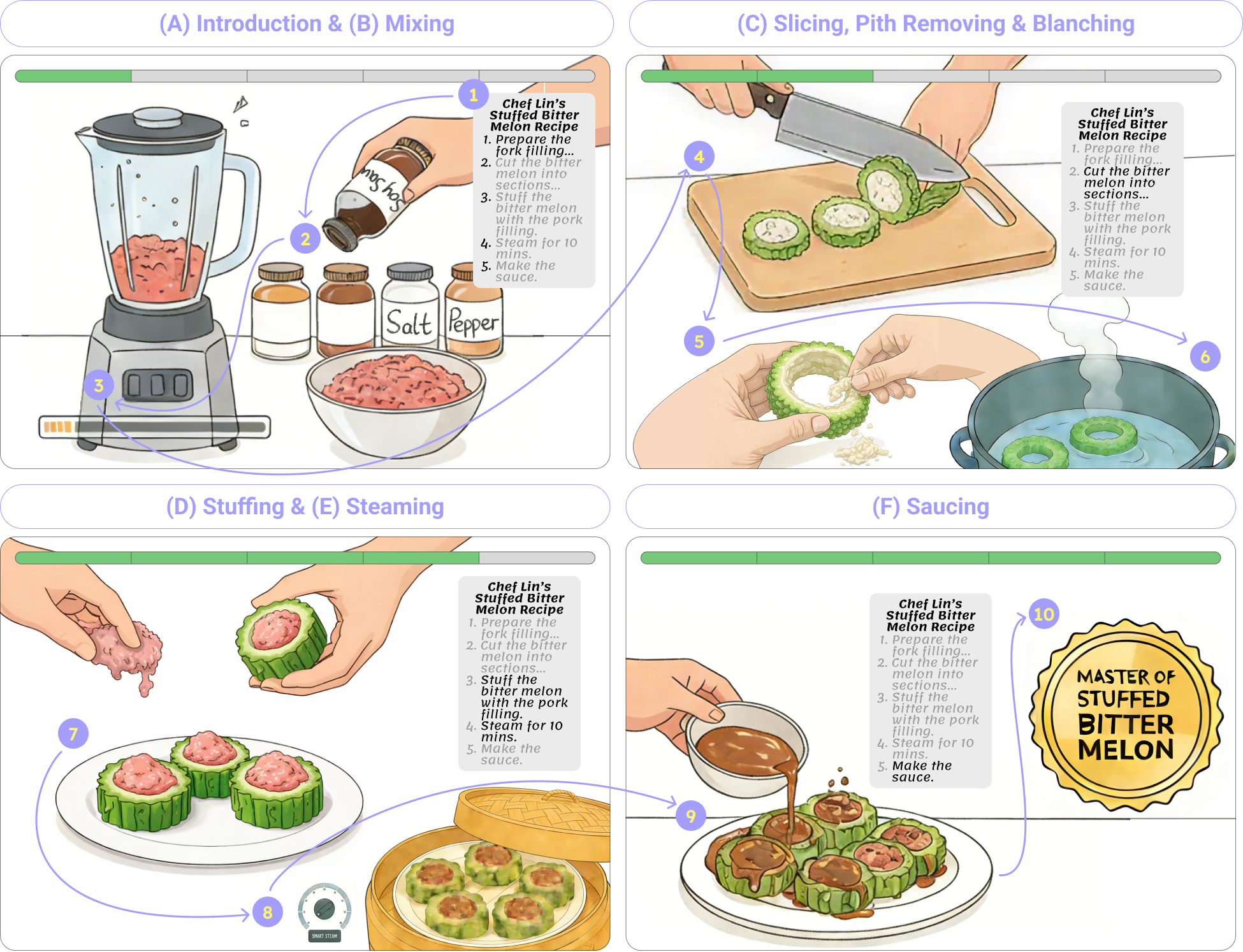}
    \caption{Player interaction sequence: (1) onboard and read the recipe; (2) choose and grab the correct seasonings and tilt them to pour into the blender; (3) press the ``Start Mixing'' button to run the blender; (4) grab a knife and cut the bitter melon into six sections; (5) remove the pith from melon rings; (6) place the rings into the pot, adjust the correct blanching timing on the timer slider and press start; (7) grab the mixed meat and stuff each melon ring until fully filled; (8) place the plate of stuffed melon rings into the steamer and steam; (9) mix the correct seasonings and pour the sauce over the steamed melon; (10) finish.} 
    \label{fig:interaction}
\end{figure}

\textit{Hakka Kitchen} is a gamified VR cooking experience designed as a scalable scaffold for ICH preservation and transmission through embodied, interactive enactment. 
Players are positioned as apprentices preparing \textit{stuffed bitter melon}, a representative Hakka dish, in a virtual kitchen environment (Figure~\ref{fig:interaction}). 
The design space for immersive ICH transmission spans from high-fidelity simulators (which prioritize procedural and sensory accuracy but supply no intrinsic motivational structure) to fully gamified experiences (which layer goals, rewards, and failure states onto an otherwise instrumental activity). 
\textit{Hakka Kitchen} contains structural features that define games, including bounded goals, immediate feedback, recoverable failure, staged progression, narrative-gated reward, and an achievement badge, which serious-games and game-based-learning research identifies as effective for sustaining persistence with material that learners may initially find effortful or unfamiliar, such as cultural heritage~\cite{habgood_motivating_2011, bellotti_serious_2012}. The game framing also enables a productive failure structure that non-game framings do not afford~\cite{kapur_productive_2008, metcalfe_memory_1994, metcalfe2017}: an unsuccessful cooking attempt becomes a failure state with narrative consequence and legible feedback, prompting re-engagement with the procedural cause, whereas in a non-game simulation the same event is simply a step performed incorrectly. This matters for culinary ICH because the cultural meaning of a constraint (e.g., pith removal as the excision of ``excess suffering'') is most legible at the moment a learner has violated it and observed the consequence, which game framings ritualize and non-game framings treat as noise.

However, \textit{Hakka Kitchen} also deliberately omits several elements that are characteristic of games in the traditional sense: there is no scoring or leaderboard, no branching narrative, no overall win-or-lose condition, and no open exploration. 
Its primary activity of cultural-procedural cooking practice is fundamentally non-game, and the game elements are layered onto that activity to support engagement rather than to constitute the experience itself. Following \citet{Deterding_Dixon_Khaled_Nacke_2011}'s definition of gamification as ``the use of game design elements in non-game contexts,'' we therefore characterize \textit{Hakka Kitchen} more precisely as a gamified VR experience: a culturally grounded cooking practice scaffolded by game elements, rather than a game in which cooking is the play activity. This distinction clarifies that the system is not asking the player to play cooking, but to enact cooking with the support of game-derived structure.

\subsection{Formative Study}

\subsubsection{Interviews with Hakka Dish Chefs}
To ground the gamified experience in authentic practice, we conducted semi-structured interviews with three Hakka chefs who have 10+ years of experience.
During the interview, we asked about 
(1) historical context of stuffed bitter melon; 
(2) ingredient preparation techniques; 
(3) standardized cooking procedures; 
and (4) cultural symbolism.
After the interviews, two researchers independently open-coded the transcripts, with a third researcher adjudicating disagreements~\cite{lazar_research_2017, Braun01012006}. Through discussion, the team consolidated the codes into a shared codebook, applied it across all three transcripts, and collectively synthesized the coded data into higher-level themes, which ultimately yielded the \textit{procedural dictionary} shown in Table~\ref{tab:dictionary}.

\subsubsection{Findings and Procedural Dictionary}

From the formative data, four tacit checkpoints emerged that chefs said novices must experience firsthand to understand the dish beyond its recipe text. These inform the \textit{procedural dictionary}: each entry records the process stage, expert insight, somatic constraint, game-mechanic implementation, and cultural connotation to inform both the interactive logic and storytelling. 

\textbf{Thickness consistency.} Chefs emphasized that rings should be ``not too thin, not too thick''; therefore, we implemented a tolerated thickness band (1--3\,cm) and surfaced it via an optional hint and corrective audio after failed attempts.

\textbf{Pith removal.} Chefs also talked about the pitfalls of working with pith. If a player attempts to blanch without scraping pith, the system blocks progression and triggers an in-world reminder explaining bitterness control.

\textbf{Blanch timing.} Chefs stressed exact timing (``one minute exactly --- too long it's mush, too short it's raw''). Players therefore choose a duration using a timer slider; incorrect timing results in a feedback state, and only after a correct attempt does Chef Lin confirm the expert target and rationale.

\textbf{Stuffing.} Chefs emphasized that underfilling causes collapse, while overfilling bursts the ring. We translated this into physics-based filling limits and feedback that reflects deformation or leakage when the threshold is violated.

We emphasized these in \textit{Hakka Kitchen} as enactable constraints alongside the cooking process, complemented by narrative cues (Chef Lin's reminders) and audio explanations after successful checkpoint actions.

\begin{table*}[t]
  \caption{The procedural dictionary: translating expert insights from the formative chef interviews into somatic constraints, game-mechanic implementations, and cultural connotations. The four checkpoint-bearing stages are shown; the mixing, steaming, and saucing stages carried no tacit checkpoint.}
  \label{tab:dictionary}
  \small
  \begin{tabular}{p{0.12\linewidth} p{0.20\linewidth} p{0.15\linewidth} p{0.25\linewidth} p{0.15\linewidth}}
    \toprule
    \textbf{Process Stage} & \textbf{Expert Insight (Translated)} & \textbf{Somatic Constraint} & \textbf{Game Mechanic Implementation} & \textbf{Cultural Connotation} \\
    \midrule
    \textbf{Slicing} & ``Cut into one-finger width\ldots not too thin or it breaks, not too thick or it stays raw.'' & \textbf{Spatial Precision:} Estimating slice thickness by eye, without a ruler. & \textbf{Collider Check:} Slices outside the 1--3\,cm tolerance band trigger a wobble animation and corrective audio; only valid slices remain interactable. & \textit{Pragmatism:} Efficiency in resource use. \\ 
    \midrule
    \textbf{Pith Removal} & ``The white pith is the source of bitterness\ldots you must scrape it clean, or it ruins the soup.'' & \textbf{Friction \& Resistance:} Scraping requires localized, repetitive force. & \textbf{Texture Erasure:} A dynamic texture mask requires 95\% removal; attempting to blanch with pith remaining blocks progression and triggers an in-world reminder on bitterness control. & \textit{Refinement:} Removing the ``excess suffering'' from life. \\ 
    \midrule
    \textbf{Blanching} & ``One minute exactly. Too long it's mush, too short it's raw\ldots add oil to keep it green.'' & \textbf{Temporal Judgment:} Judging elapsed time without a visible clock. & \textbf{Withheld-Parameter Timer:} Players set a duration without being told the target; incorrect timing triggers error feedback and blocks progression; after a correct attempt, Chef Lin confirms the expert target and rationale. & \textit{Transformation:} The moment of change. \\ 
    \midrule
    \textbf{Stuffing} & ``Stir clockwise to make the meat `Q' (bouncy)\ldots stuff it full but don't break the ring.'' & \textbf{Rhythmic Consistency \& Fragility:} Balancing force against material limits. & \textbf{Gesture Detection \& Destructibility:} Physics-based filling limits; an underfilled ring collapses and an overfilled ring bursts, with visible deformation and leakage feedback. & \textit{Resilience \& Reunion:} ``\textit{Ku jin gan lai}''---bitterness ends, sweetness begins; the filled ring signifies family reunion (\textit{tuanyuan}). \\ 
    \bottomrule
  \end{tabular}
\end{table*}

\subsection{Experience Design}
\subsubsection{Design Rationale}
\label{sec:rationale}
To determine the optimal medium for transmitting process-based culinary ICH, we evaluated interface modalities based on their ability for isomorphic mapping~\cite{macaranas_what_2015}. We first ruled out desktop and mobile interfaces because they rely on symbolic input that abstracts the culinary process~\cite{ishii_tangible_1997, klemmer_how_2006}.
During initial prototyping, we tested standard 6-DoF controllers (thumbstick-and-button input). 
For direct-touch cooking actions, handheld, lightweight controllers introduce a tool-mediated layer (e.g., button presses on controllers) that can feel mismatched with the intended gesture and can disconnect users from direct physical practice, thereby increasing cognitive load and potentially hindering deeper cognitive absorption~\cite{Palombo_Weber_Wyszynski_Niehaves_2024}.

We therefore adopted controller-free hand tracking with a visible hand avatar to better support reality-based interaction and more isomorphic action--effect mapping for cooking behaviors that are performed directly by hand in real kitchens---for example, removing pith and stuffing the meat filling---consistent with evidence that bare-hand input is more intuitive for direct-touch actions~\cite{Luong_Cheng_Moebus_Fender_Holz_2023}.
This also aligns with evidence that direct manipulation engages natural motor schemas, facilitating stronger sensorimotor memory traces, engagement, and cognitive absorption in complex procedural tasks~\cite{Saran_2025, Pangestu_Primasari_Sidhi_Wibisono_Setyohadi_2022, Johnson-Glenberg_2018}.

\subsubsection{Design Goals}
We organized the experience around five cooking stages derived from the expert interviews, and distilled the dictionary's checkpoints into four design goals, drawing on preliminary findings and a literature review. 

\textbf{DG1: Embodied enactment via natural hand interaction.}
Following the prototyping insights in Section~\ref{sec:rationale}, players complete each stage through direct hand actions.
These interactions are designed to align with embodied real-world actions, so that procedural constraints (e.g., sequencing prerequisites, quantity tolerance, timing windows) are learned through practice, creating a sense of presence that reinforces motor memory and procedural flow.

\textbf{DG2: Sensory and immediate feedback.} 
The formative study revealed that sensory feedback, particularly visual state changes such as the color change of bitter melon that indicates its readiness, is important in the process. Because VR cannot reproduce full sensory cues, such as smells, we also add auditory cues to make progress and constraint violations legible, such as the slice sound when the knife cuts through the melon. 
This creates a closed action--feedback loop where players act, perceive consequences, and adjust, reinforcing embodied cognition through enactment.

\textbf{DG3: Intrinsic narrative integration.} 

We aim for players across cultural backgrounds and cooking-experience levels to learn both the cooking process and the cultural meanings behind this ICH dish. Therefore, we adopted the situated-learning framework~\cite{lave_situated_1991} to contextualize the somatic struggle, framing players as apprentices in the kitchen. We explicitly rejected a ``Family Role'' to avoid dissonance with players' personal family histories. Instead, the ``apprentice'' role provides a universal, culturally valid reason for the user's initial incompetence and gradual mastery. Chef Lin, as the mentor, provides cooking guidance through voice-over to keep players motivated by offering clear goals.
Chef Lin's narration is also presented step by step and tied to the just-completed action, so that cultural meaning is encountered in situ rather than as detachable exposition. For example, narration following blanching links timing discipline to a personal anecdote, while the stuffing step links the dish's bitterness-to-sweetness metaphor to resilience.
This narrative design also grounds the story in Chef Lin's personal experiences and presents it in the first person; we aimed to make the narration more relatable and engaging for players.

\textbf{DG4: Lightweight instructions and productive hints.}
To avoid overwhelming players, especially for kitchen novices, we aim to provide only lightweight procedural instructions for the next goal without prescribing every micro-action, leaving room for exploration.
Immediate feedback distinguishes correct from incorrect actions, and targeted reminders surface interview-derived pitfalls. For example, if a melon is underfilled or if the pith has not been removed before blanching, an audio cue is triggered to draw attention to the error. 
Drawing from Metcalfe's Error-Based Learning Theory~\cite{metcalfe2017}, to leverage learning from mistakes, we deliberately withhold certain key values emphasized in the formative study by chefs, so that players must test and correct. For example, during the blanching stage, players are presented with a timer slider but are not told the exact timing. Therefore, players must experiment across several attempts to find the correct timing; incorrect attempts trigger error feedback and block progression to the next stage. 
Hints provide additional tolerances and rationales to reinforce the checkpoints in Table~\ref{tab:dictionary}, such as slicing thickness.

\subsubsection{Engagement Strategies}
Achievement systems can serve as strong motivators in serious games~\cite{bellotti_serious_2012, habgood_motivating_2011}. To sustain player motivation and ensure continuity across the cooking sequence, we integrated several engagement strategies that operate at both procedural and cultural levels.

\textbf{Progress indicator.} A visual progress bar tracks advancement through the five cooking stages, conveying the structured, sequential nature of preparing the dish. 
\textbf{Bonus narratives.} Each completed stage unlocks a new segment of Chef Lin's narration, embedding cultural knowledge into the flow of the experience, incrementally revealing the stories behind this ICH dish. 

\textbf{Achievement badge.} Upon completing all five stages, players earn a ``Hakka Kitchen'' badge that certifies their completion of the stuffed bitter melon procedure. 
This acts as a tangible symbol of recognition and accomplishment, sustaining engagement through to the end of \textit{Hakka Kitchen}.

\subsection{Experience Flow}
\label{sec:gameflow}

\begin{figure}[htbp]
    \centering
    \includegraphics[width=0.99\linewidth]{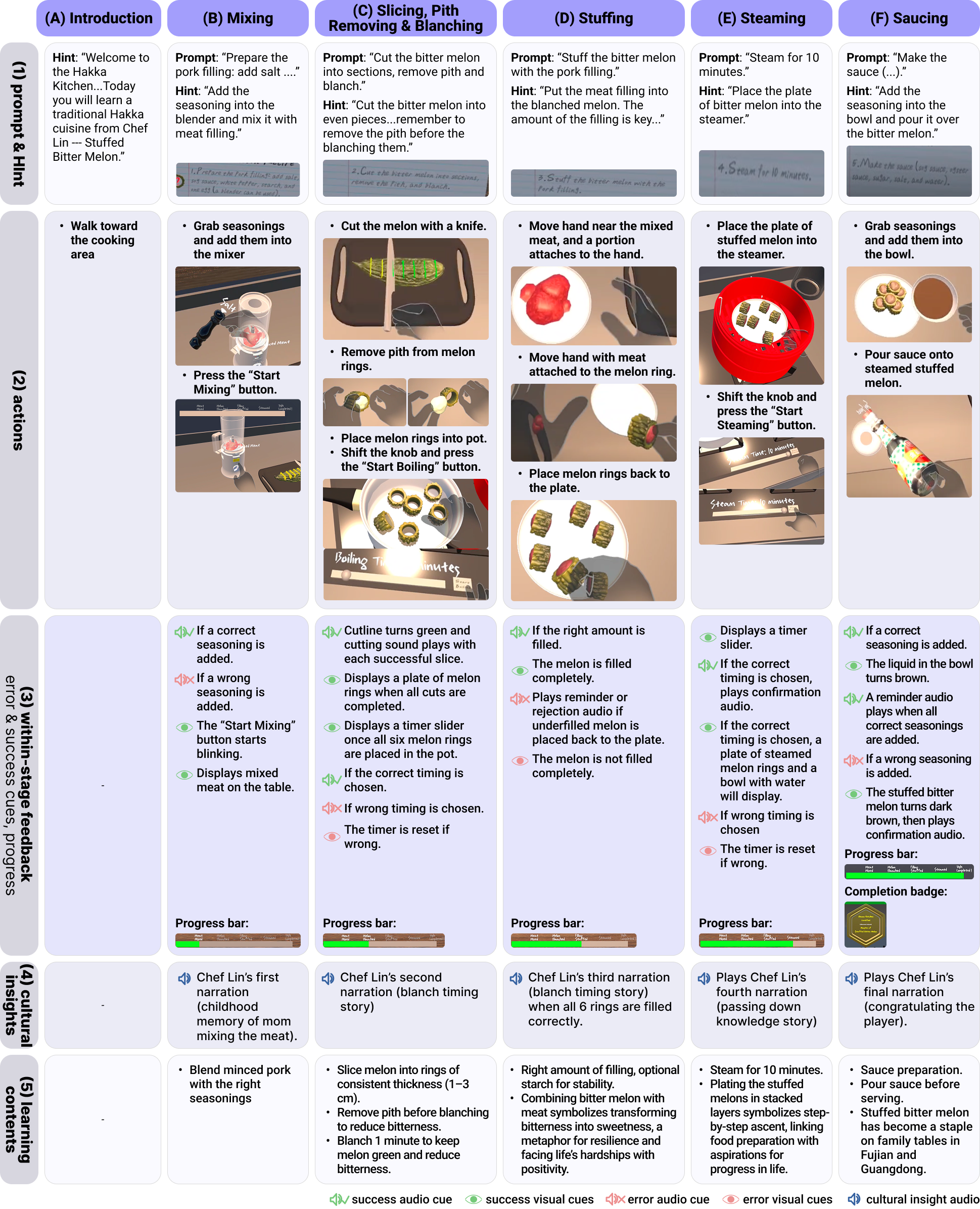}
    \caption{Experience Flow. Players progress through the stages of (A) introduction, (B) mixing, (C) slicing, pith removal and blanching, (D) stuffing, (E) steaming, and (F) saucing. Each stage presents players with (1) prompts and hints, (2) actions, (3) within-stage feedback and progress indicators, and (4) after-stage cultural-insight bonuses, and (5) the learning contents each stage conveys, illustrating how procedural steps, feedback, and cultural narratives are integrated across the experience.}
    \label{fig:gameflow}
\end{figure}

Upon entering \textit{Hakka Kitchen}, players are onboarded as apprentices to Chef Lin, establishing the framework (Figure~\ref{fig:gameflow}A).
The experience unfolds over five distinct culinary stages derived from the expert interviews: (B) Mixing, (C) Slicing, Pith Removal \& Blanching, (D) Stuffing, (E) Steaming, and (F) Saucing (Figure~\ref{fig:gameflow}).
Each cooking stage is accompanied by (1) prompts \& hints, (2) actions, (3) within-stage feedback, and (4) after-stage cultural insight bonuses.
For instance, prompts are presented on the recipe sheet, and hints are complemented by a deliberately open-ended directive from Chef Lin. For example, for (B) Mixing, the prompt is ``prepare the pork filling'' (Figure~\ref{fig:gameflow}(1)).
Actions for (C) Slicing \& Pith Removal require players to attempt the task using physics-based hand gestures while the system checks the somatic constraints (e.g., checking if slice thickness is within the 1--3\,cm valid range); hints such as ``Prepare the melon, but be gentle'' set the goal without specifying the technique (Figure~\ref{fig:gameflow}(2)).
After completion of an action, the system cues success or failure: if a constraint is violated (e.g., wrong blanching time), audio-visual feedback prompts a retry; on success, a confirmation chime plays and the ingredient's visual state updates (e.g., meat turns sticky) (Figure~\ref{fig:gameflow}(3)). 
Meanwhile, the progress bar reflects the current progress and state of success.
Upon successful completion of a stage, the system unlocks the bonus insights where Chef Lin explains the deeper meaning of the action (e.g., ``The filled ring stands for \textit{tuanyuan} --- the family whole and together'') (Figure~\ref{fig:gameflow}(4)).
After completion of the whole experience, players receive a badge.

\subsubsection{Matched Video Condition}
\label{sec:video}
To isolate the effect of embodied enactment (rather than content exposure), we created a matched control condition in which participants watched a video in VR presenting the same five-stage procedural content and cultural narration.
We stitched multiple recipe clips into a continuous sequence, overlaid the narration, and when available, prioritized a chef's-eye perspective to align the viewpoint.
The video was presented in VR with standard playback control (e.g., pause or seek), but without the ability to manipulate ingredients or trigger procedural consequences.

\subsection{Implementation}
\begin{figure}[htbp]
    \centering
    \includegraphics[width=\linewidth]{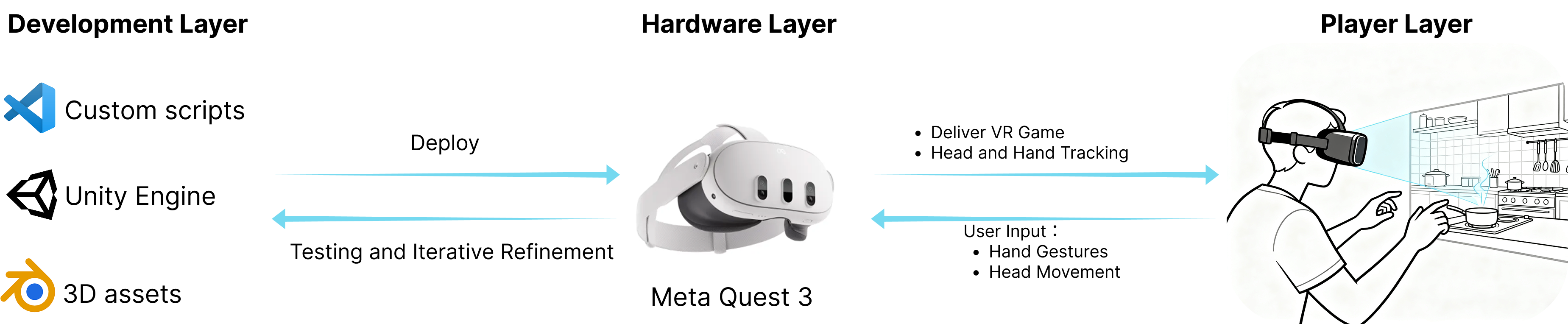}
    \caption{System architecture. The VR gamified experience is developed in Unity and deployed to the Meta Quest 3, with iterative testing feeding results back into the development environment. On the hardware, Quest 3 delivers the gamified experience to the player while capturing head and hand inputs, which are processed and returned through the system to sustain real-time interaction.}
    \label{fig:system}
\end{figure}

We implemented \textit{Hakka Kitchen} in Unity 6\footnote{\url{https://docs.unity.com}} and deployed it on Meta Quest 3\footnote{\url{https://www.meta.com/quest/quest-3/}}.
The system processes head and hand tracking, resolves physics interactions (e.g., grabbing, cutting, scooping, stuffing), triggers state-dependent feedback (visual and audio cues, constraint checks, reminders), and unlocks narration based on stage completion.
Figure~\ref{fig:system} summarizes the runtime loop from user input to simulation update and feedback rendering.

\section{Methods}\label{sec:Methods}
\subsection{Study Design}
We conducted a two-arm, between-subjects experiment comparing an \emph{interactive gamified VR cooking experience} to a control \emph{condition in which participants watched a video within the same VR environment} to isolate the effect of embodied enactment.
% Why between-subjects
We chose a between-subjects rather than within-subjects design because first exposure to making the recipe is non-recoverable across conditions; therefore, a between-subjects study reduces the carried-over context effect and range effect that appear in a substantial fraction of within-subjects studies and are not fully neutralizable by counterbalancing~\cite{poulton_unwanted_1973, greenwald_within-subjects_1976, charness_experimental_2012}.
In the gamified VR condition, participants completed the full preparation workflow using controller-free hand tracking and physics-based object manipulation with in-game guidance, cues, and hints. 
In the video control condition, participants watched a first-person video of the same workflow (see Section~\ref{sec:video}) within the VR environment with standard playback controls but without the ability for other actions such as manipulating ingredients. 
This arrangement ensures that both groups experienced the same immersive environment and content, while differing only in whether participants physically enacted or passively observed, which is the comparison logic established for isolating instructional features in immersive-learning research~\cite{makransky2021cognitive, makransky_benefits_2022}.

\subsection{Participants}
As shown in Table~\ref{tab:demographics-summary} (details in Appendix~\ref{Demogrpahic}), we recruited 40 participants (ages 18--37) from the university's mailing lists.
Participants came from diverse regional backgrounds across China, providing variation within a young, educated sample, and were cultural novices with respect to Hakka culinary ICH.
We screened for low prior familiarity with Hakka cuisine. 
China’s culinary landscape is highly regionalized, while Hakka cuisine is a specific subcultural tradition concentrated primarily in parts of Fujian and Guangdong and maintained through community-specific transmission.
Therefore, participants recruited outside these communities had little prior exposure with the cooking procedure or the cultural meanings embedded in Hakka dishes; we confirmed low familiarity during screening.
Eligibility required normal or corrected-to-normal vision and no history of severe motion sickness or vestibular disorders. 
Before the session, participants provided demographic information (age, gender identity, highest level of education, cultural background, weekly gaming hours, and self-reported cooking experience). 
All participants gave informed consent and received 15 CNY as compensation.

\begin{table}
    \centering
    \caption{Participant demographics by condition. Values are $M$ ($SD$) or counts.}
    \label{tab:demographics-summary}
    \resizebox{\textwidth}{!}{
    \begin{tabular}{l c c cc ccc cccc ccc}
    \toprule
    & & & \multicolumn{2}{c}{Gender} & \multicolumn{3}{c}{Education} & \multicolumn{4}{c}{Weekly gaming (h)} & \multicolumn{3}{c}{Self-report $M$ ($SD$)} \\
    \cmidrule(lr){4-5}\cmidrule(lr){6-8}\cmidrule(lr){9-12}\cmidrule(lr){13-15}
    Condition & $n$ & Age $M$ ($SD$) & F & M & Bach. & Mast. & HS & 0 & 0--3 & 3--7 & $\geq$7 & Cooking & Prior VR & Motion \\
    \midrule
    Gamified VR (enactment) & 20 & 26.4 (3.4) & 13 & 7 & 8  & 12 & 0 & 4 & 11 & 2 & 3 & 3.4 (1.8) & 2.4 (1.5) & 2.3 (1.6) \\
    VR video (observation)  & 20 & 22.7 (2.0) & 12 & 8 & 13 & 6  & 1 & 3 & 8  & 4 & 5 & 3.5 (1.9) & 2.7 (1.5) & 2.7 (1.8) \\
    \bottomrule
  \end{tabular}}
\end{table}

\subsection{Procedure}
\subsubsection{Session Flow}
\begin{figure}
  \centering
  \includegraphics[width=\linewidth]{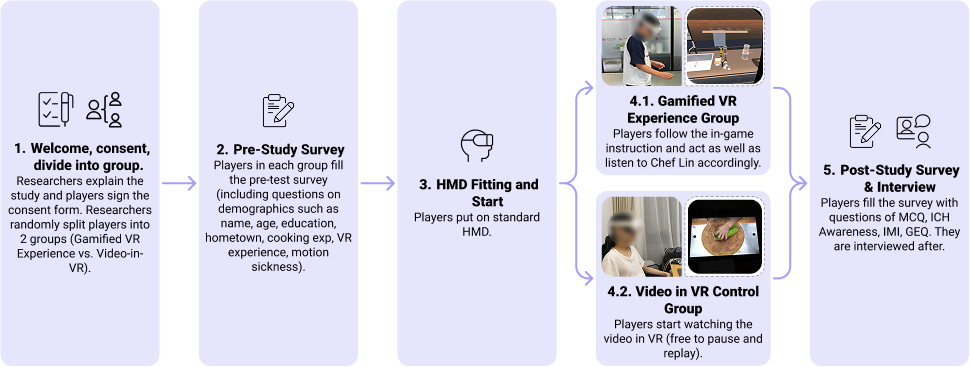}
  \caption{Study procedure: (1) welcome and consent; (2) random group assignment and pre-study survey; (3) HMD fitting; (4) experience; (5) post-study survey and interview.}
  \label{fig:flow}
\end{figure}

Each session lasted approximately 30 minutes and followed the same structure (Figure~\ref{fig:flow}): 
(1) Participants reviewed the information sheet and provided written informed consent; 
(2) Researchers randomly assigned participants to the experimental or control group, then participants completed the pre-study surveys;
(3) Participants put on the head-mounted display (HMD), verified comfort, and had the safety boundaries explained to them;
(4) Participants experienced one condition (either the gamified VR experience or the matched VR video) in a single sitting;
(5) After reporting completion, participants removed the headset and completed the post-study survey, after which researchers conducted a 5- to 10-minute semi-structured interview.

\subsubsection{Measurements}
We measured the following items in the post-study survey:

\textbf{Procedural knowledge instruments.} The Multiple-Choice Quiz (MCQ) assessed critical parameter knowledge with six questions, 
focused on process, parameters, and dependencies emphasized in the design (e.g., tolerance constraints such as thickness).

\textbf{Interest and motivation.} 
We used the Interest/Enjoyment and Perceived Competence subscales of the Intrinsic Motivation Inventory (IMI) short form to assess intrinsic motivation immediately post-exposure~\cite{mcauley_psychometric_1989}.

\textbf{Game Experience.} 
The Game Experience Questionnaire (GEQ) captures immersion (Sensory \& Imaginative Immersion), flow, competence, positive and negative affect, challenge, and tension~\cite{ijsselsteijn_game_2013}.

\textbf{Cultural heritage awareness.} 
To index perceived value, interest, and attitudes toward ICH, we used a short form adapted from the ICH Scale~\cite{liu_measuring_2024}. This instrument indexes perceived transmission, localization, vitality, and association in cultural practices.

\subsubsection{Interview} 
We conducted semi-structured interviews focusing on (1) perceived agency and pacing, (2) learnability and clarity of procedural constraints, (3) perceived cultural salience (what felt culturally meaningful and why), (4) realism boundaries (what felt missing from real cooking), and (5) design recommendations.
Interviews were audio-recorded and transcribed.

\subsection{Analysis}
We adopted a mixed-methods analytic plan with $\alpha=.05$; we report effect sizes and 95\% confidence intervals for all key comparisons.

\textbf{Quantitative data.} 
All quantitative data were analyzed in R. Questionnaire responses were analyzed at the dimension level. Because responses were Likert-type and violated normality (Shapiro--Wilk, $p$ < .05), between-group comparisons used two-sided Mann--Whitney $U$ tests (equivalent to Wilcoxon rank-sum), reported with Cliff's δ and bootstrapped 95\% CIs (10{,}000 resamples). Given eight questionnaire comparisons, we report uncorrected $p$-values and indicate which effects survive Holm correction. Procedural knowledge was scored as a six-item sum (0--6) and compared with the same test; per-item accuracy was compared with Fisher's exact tests, appropriate given expected cell counts below five.

\textbf{Qualitative data.} 
To analyze the post-session semi-structured interviews, we employed a two-researcher codebook thematic analysis~\cite{braun_one_2021} operationalized through the Framework Method~\cite{gale_using_2013}, which pairs the systematic, comparison-oriented charting of Framework analysis with structured, codebook-based coding.
The process was iterative.
Two researchers independently open-coded a balanced 30\% subset of transcripts (balanced across conditions) to draft an initial codebook of inductively derived codes, then discussed divergences to refine code definitions and boundaries and to reach a shared coding frame~\cite{mcdonald_reliability_2019}.
The codebook was then applied to the full corpus, but coding and the codebook were revised in repeated passes: adding, merging, and re-specifying codes as new transcripts surfaced cases that the current codebook did not fit.
The researchers iterated between coding and theme development, clustering and re-clustering charted codes into candidate themes. 
We continued this cycle until the codebook stabilized and additional transcripts yielded no new codes or themes, and the team judged the theme structure to coherently account for the data.
The full codebook is reported in Appendix~\ref{Codeboook}.

\section{Results}\label{sec:Results}

We report quantitative outcomes comparing the gamified VR experience (Experiment) and video in VR (Control) conditions (RQ3), followed by qualitative themes from our codebook thematic analysis (RQ2).

\subsection{RQ3: How enactment and observation differ in transmitting layers of culinary heritage}

We first report descriptive statistics and Shapiro--Wilk normality checks. Given non-normality for several outcomes, all between-condition comparisons use the two-sided Mann--Whitney $U$ test (equivalent to the Wilcoxon rank-sum test), reported with Cliff's $\delta$ and bootstrapped 95\%\ confidence intervals (Table~\ref{tab:desc_normality_infer}). Because we conduct eight questionnaire comparisons, we report uncorrected $p$-values alongside effect sizes and CIs and, for transparency, indicate which outcomes additionally survive Holm correction across the eight tests ($\dagger$); we anchor our interpretation on these robust effects. We describe distributions using medians and IQRs, but note that the Mann--Whitney $U$ test evaluates whether one condition tends to yield higher ranks overall rather than testing a difference in medians per se.

\begin{table*}
\centering
\caption{Descriptive statistics, Shapiro--Wilk normality checks, and between-condition comparisons (gamified VR vs.\ video in VR). Values are $M$ ($SD$); medians and IQRs are reported in text. Normality columns give the Shapiro--Wilk $W$ and $p$ within each condition ($n=20$ per group). Between-condition tests are two-sided Mann--Whitney $U$ with Cliff's $\delta$ and bootstrapped 95\%\ CIs (10{,}000 resamples). Procedural Knowledge is a six-item sum score ($0$--$6$). Bold rows are significant at the uncorrected $p<.05$ level; $\dagger$ marks outcomes that remain significant after Holm correction across the eight questionnaire outcomes.}

\label{tab:desc_normality_infer}
\resizebox{\textwidth}{!}{
\begin{tabular}{lcc|cc|c|cc|ccc}
\toprule
& \multicolumn{2}{c}{$M$ ($SD$)}
& \multicolumn{2}{c}{Shapiro--Wilk $W$}
& \multicolumn{2}{c}{Normality $p$}
& \multicolumn{3}{c}{Between-condition test} \\
\cmidrule(lr){2-3}\cmidrule(lr){4-5}\cmidrule(lr){6-7}\cmidrule(lr){8-10}
Outcome & Gamified & Video & Gamified & Video & Gamified & Video & $U$ & $p$ & $\delta$ [95\%\ CI] \\
\midrule
\textbf{IMI: Interest/Enjoyment}$\dagger$   & 5.66 (0.85) & 4.30 (0.59) & .955 & .954 & .451 & .436 & 361.5 & \textbf{$<$.001} & 0.81 [0.60, 0.95] \\
IMI: Perceived Competence                   & 4.65 (1.19) & 4.27 (0.54) & .960 & .964 & .539 & .628 & 262.5 & .092 & 0.31 [$-$0.04, 0.66] \\
\textbf{GEQ: Sensory \& Imaginative}$\dagger$& 5.66 (0.91) & 4.62 (0.85) & .924 & .967 & .118 & .701 & 325.0 & \textbf{$<$.001} & 0.62 [0.33, 0.86] \\
\textbf{GEQ: Positive Affect}               & 5.75 (0.94) & 4.90 (0.92) & .936 & .893 & .203 & .030 & 292.5 & \textbf{.013} & 0.46 [0.12, 0.76] \\
\textbf{Awareness: Transmission}$\dagger$    & 5.70 (1.03) & 4.00 (1.26) & .878 & .887 & .017 & .024 & 339.5 & \textbf{$<$.001} & 0.70 [0.46, 0.89] \\
\textbf{Awareness: Vitality}                & 5.60 (1.10) & 4.45 (1.47) & .883 & .925 & .020 & .126 & 289.5 & \textbf{.013} & 0.45 [0.12, 0.73] \\
Awareness: Association                       & 5.65 (1.35) & 4.90 (1.92) & .816 & .885 & .001 & .022 & 243.5 & .231 & 0.22 [$-$0.14, 0.56] \\
Awareness: Localization                      & 4.80 (1.32) & 5.20 (1.61) & .892 & .863 & .029 & .009 & 155.5 & .218 & $-$0.22 [$-$0.56, 0.13] \\
Procedural Knowledge (0--6)                  & 5.15 (0.88) & 5.05 (0.76) & .826 & .816 & .002 & .002 & 219.5 & .582 & 0.10 [$-$0.24, 0.43] \\
\bottomrule
\end{tabular}}
\end{table*}

\subsubsection{Interest \& Motivation (IMI)}
As shown in Figure~\ref{Group}, participants in the gamified VR experience group reported higher Interest/Enjoyment ($Mdn=5.7$, IQR $[4.95, 6.45]$) than the video group ($Mdn=4.4$, IQR $[3.95, 4.45]$), a significant difference (Mann--Whitney $U=361.5$, $p<.001$, $\delta=0.81$). Perceived Competence did not differ significantly (gamified $Mdn=4.7$, IQR $[4.30, 5.20]$; video $Mdn=4.4$, IQR $[3.95, 4.60]$; $U=262.5$, $p=.092$).
The increase in interest reflects the motivational pull of interactivity and narrative-driven tasks, whereas the absence of a competence difference suggests both modalities offered comparable challenge and task clarity. The heightened immersion and positive affect are consistent with qualitative reports of first-person, close-up, controlled stepping in the gamified VR experience; the non-significant competence difference is consistent with our Discussion account that a single exposure without multisensory cues such as heat and smell makes stable self-efficacy difficult to form.

\subsubsection{Game Experience (GEQ)}
As shown in Figure~\ref{Group}, the gamified VR experience group scored higher than the video group on Sensory \& Imaginative Immersion (gamified $Mdn=5.8$, IQR $[5.00, 6.33]$; video $Mdn=4.7$, IQR $[4.12, 5.04]$; $U=325.0$, $p<.001$, $\delta=0.62$) and on Positive Affect (gamified $Mdn=5.9$, IQR $[4.95, 6.50]$; video $Mdn=5.1$, IQR $[4.60, 5.60]$; $U=292.5$, $p=.013$, $\delta=0.46$). The advantage in Sensory \& Imaginative Immersion indicates stronger perceptual immersion under enactment, and the higher Positive Affect reflects a more emotionally rewarding experience. Under Holm correction, the Sensory \& Imaginative effect remains significant, while Positive Affect becomes borderline ($p_{\text{Holm}}=.063$); we therefore treat the affect result as suggestive.

\subsubsection{Cultural-Heritage Awareness}
\label{localization}
As shown in Figure~\ref{Group}, three of the four awareness sub-dimensions favored the gamified VR experience group. Transmission was higher for the gamified group (gamified $Mdn=6.0$, IQR $[5.00, 6.25]$; video $Mdn=4.0$, IQR $[3.75, 5.00]$; $U=339.5$, $p<.001$, $\delta=0.70$), as was Vitality (gamified $Mdn=5.5$, IQR $[5.00, 6.25]$; video $Mdn=5.0$, IQR $[3.75, 5.25]$; $U=289.5$, $p=.013$, $\delta=0.45$). Association did not differ significantly (gamified $Mdn=6.0$, IQR $[5.00, 6.25]$; video $Mdn=5.0$, IQR $[4.00, 6.25]$; $U=243.5$, $p=.231$).
Localization was the one dimension on which the video group scored numerically higher, though the difference was not significant (gamified $Mdn=5.0$, IQR $[4.00, 6.00]$; video $Mdn=5.5$, IQR $[5.00, 6.00]$; $U=155.5$, $p=.218$, $\delta=-0.22$). The non-significant Association and Localization results suggest both modalities were comparably (in)effective at conveying communal connection and situating the practice in its local context. The Localization pattern is consistent with our later finding that manual load can crowd out relational and place-based narration during enactment (Section~\ref{sec:heritage_recognition}, action masking); embedding localization cues at additional, lower-load steps is therefore a direction for closing this gap rather than a benefit the current design already delivers.

\begin{figure}[h]
    \centering
    \includegraphics[width=\linewidth]{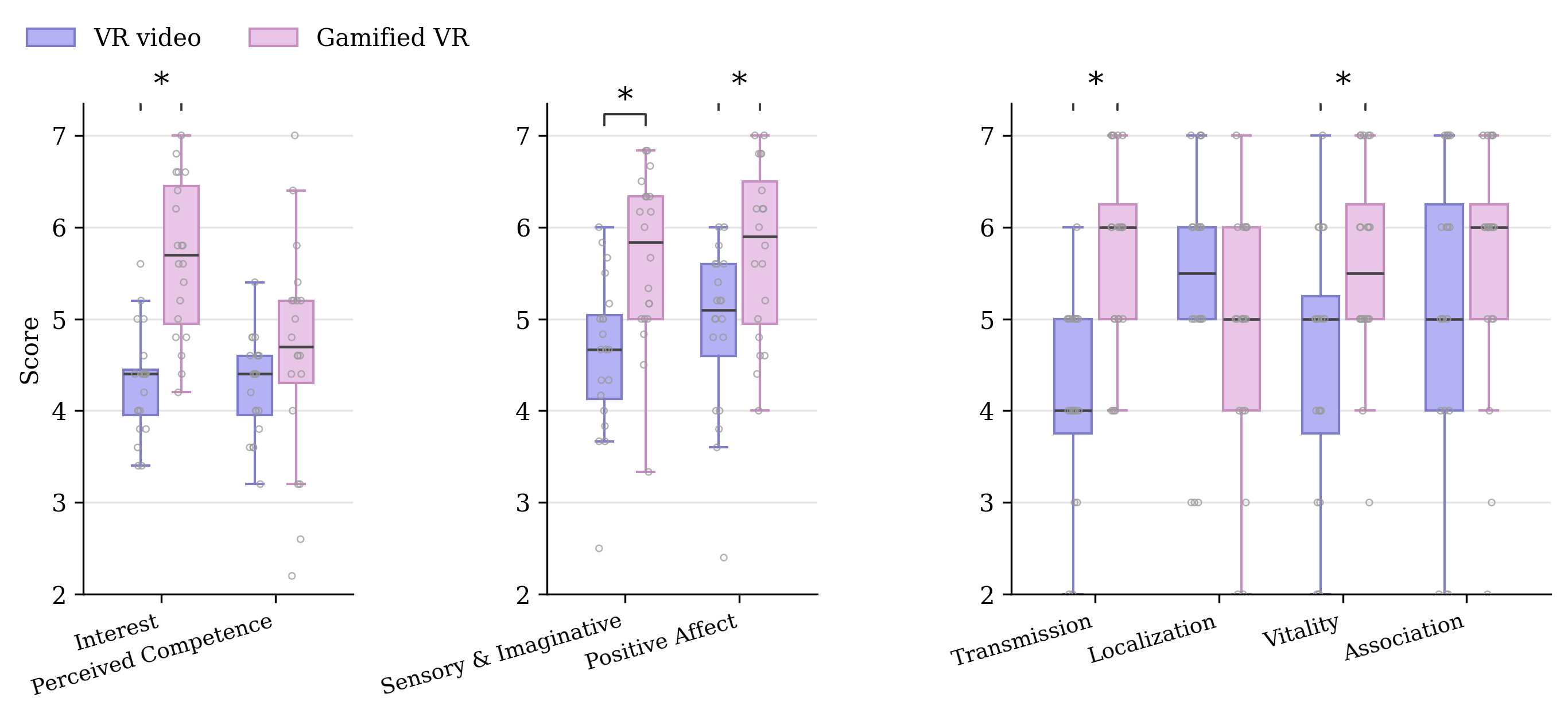}
    \caption{Intrinsic Motivation Inventory (IMI), Game Experience Questionnaire (GEQ), and Cultural-Heritage Awareness scores by condition. Asterisks mark uncorrected significance ($*\,p<.05$; see Table~\ref{tab:desc_normality_infer} for exact $p$, effect sizes, and Holm-corrected outcomes).}
    \label{Group}
    \label{Group}
\end{figure}

\begin{figure}[h]
    \centering
      \includegraphics[width=0.9\linewidth]{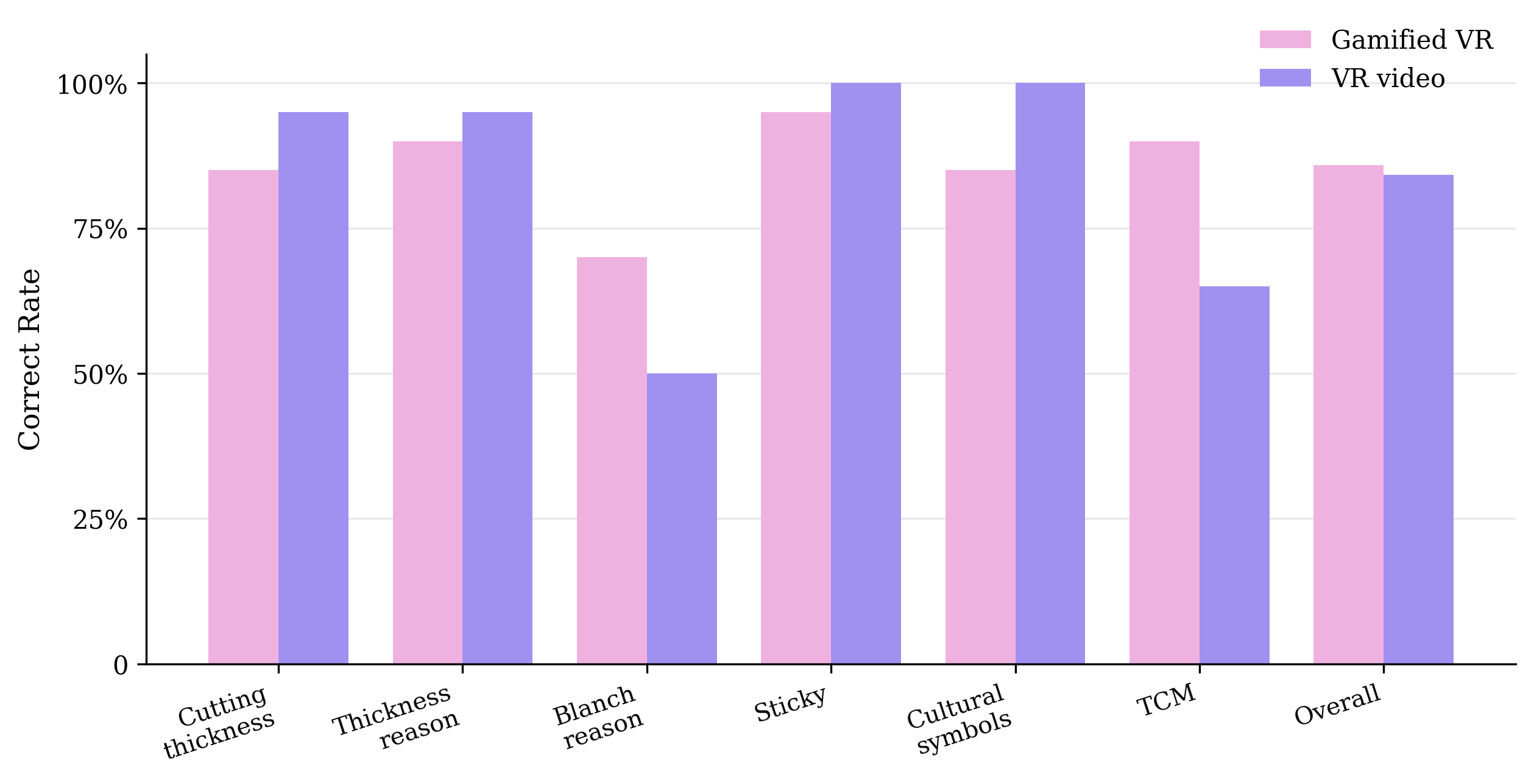}
      \caption{Per-item MCQ accuracy by condition; no comparison reached significance (Fisher's exact, $p$ > .05).}
    \label{ccr}
\end{figure}

\subsubsection{Procedural Knowledge Quiz}
\label{quiz}
\label{quiz}
As shown in Figure~\ref{ccr}, procedural knowledge did not differ between conditions. Scored as a six-item sum ($0$--$6$), the two groups performed almost identically (gamified $M=5.15$, $Mdn=5.0$; video $M=5.05$, $Mdn=5.0$; Mann--Whitney $U=219.5$, $p=.582$, $\delta=0.10$). Per item, Fisher's exact tests---appropriate here because expected cell counts fall below five---showed no significant differences (all $p>.12$); the largest gaps, both non-significant, were the blanching-purpose item (gamified 14/20 vs.\ video 10/20, $p=.333$) and the bitter-melon rationale item (18/20 vs.\ 13/20, $p=.127$).
In the interviews, participants in both conditions described the resumable ``mistake--reset--redo'' loop as a memory anchor and frequently credited explicit cues---parameter ranges such as slice thickness and stuffing tightness---with aiding recall and transfer. This may explain the null quiz result: recognition-level knowledge transferred under both modalities, whereas real-world operational or parametric performance, which we did not measure, may be more sensitive to enactment.

\subsection{RQ2: Embodied Doing}

\subsubsection{Knowing-About vs.\ Knowing-How: Enactment Converts Procedure into Consequential Felt Knowledge}
\label{sec:knowing-about}

Participants in both groups converged on a distinction between learning what steps to take and learning what those steps feel like.
As shown in Figure~\ref{ccr}, both groups performed comparably on the Procedural Knowledge Quiz, and the video-in-VR participants noted the clarity of the recipe sequence: \textit{``the order from the VR tutorial is clear. [You] just follow step by step''} (C3). 
However, the Video-in-VR participants stated directly the limit of watching: \textit{``watching videos is like scrolling food bloggers that the eyes learn but the hands don't''} (C2). 
Especially when ``appropriate amount'' was mentioned in the recipe for both conditions, Video-in-VR participants reported being left without a referent: \textit{``I had no idea how much''} (C9) or \textit{``how much in practice is anyone's guess''} (C10). 
Even participants with more cooking experience, such as C3 and C20, reflected that watching makes cooking look simple in a way that overpromises against real execution.

\begin{figure}[h]
  \centering
  \includegraphics[width=\linewidth]{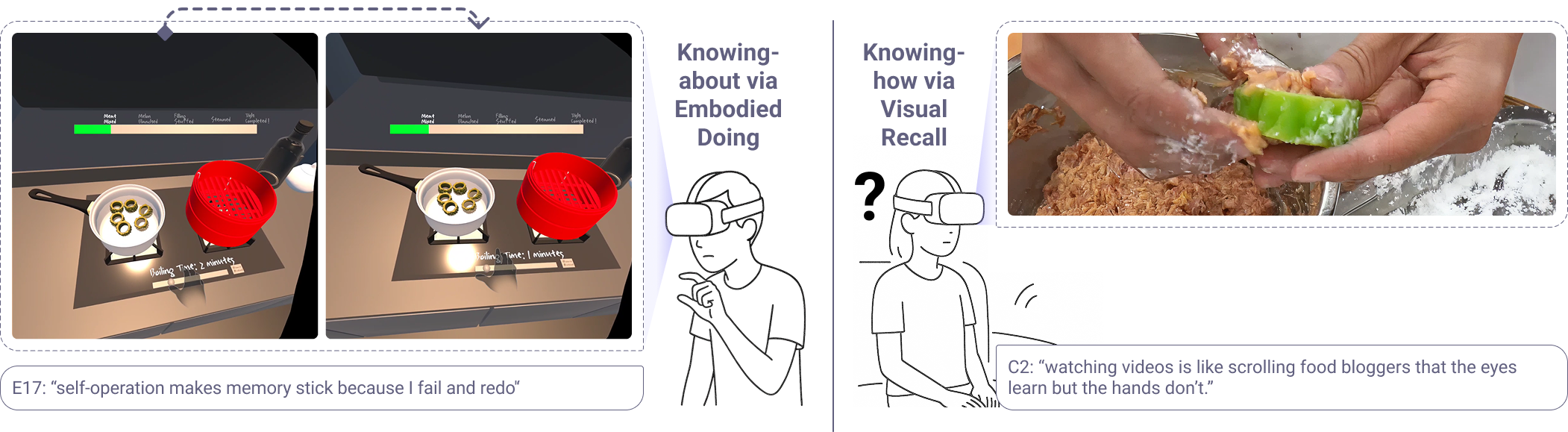}
  \caption{Knowing-how via embodied doing vs. knowing-about via visual recall. The gamified VR experience fosters a recoverable error--repair loop where physical mistakes (e.g., wrong timing for blanching) become somatic memory anchors. In contrast, the video condition creates a transfer gap, where, for example, participants retain the action for filling but lack the tacit confidence to execute the task.}
  \label{fig:embody}
\end{figure}

On the contrary, in the gamified VR experience group, the same tacit parameters became bodily-legible through enactment. E1 stated that \textit{``through doing you get a direct impression of width, not just a flat `1--3 cm' on paper.''} 
Because participants can actually practice by acting and moving within the virtual kitchen environment, they felt \textit{``acting myself''} (E4), such as \textit{``laying out materials on the kitchen table and following the kitchen circulation route helps me memorize better''} (E13). 
This shows that participants can see the parameters through video, but they can only feel a sense of these parameters through doing. 
Moreover, consequential errors did similar work; E17 expressed \textit{``self-operation makes memory stick because I fail and redo''}, and E1 stated they memorized to fill the bitter melon more and use the correct seasoning because of the error hints triggered when they made mistakes while doing. 
Video-in-VR group participants mentioned that they could imagine that they would make mistakes by imagination, such as \textit{``I didn't make any errors, I was only watching, but I can imagine over-stuffing''} (C1). 
These highlight that enactment converts abstract procedure into encoded experience through tangible, recoverable failure.
See Figure~\ref{fig:embody}.

\subsubsection{Agency and Pacing: Embodied Doing Enabled Micro-Control but Exposed Macro-Constraint and Physics-Simulation Breakdowns}
\label{sec:agency}
 
Enactment in the gamified VR group produced an experience that participants consistently split between micro-agency gained and macro-constraint exposed. 
At a micro level, such as gesturing, the gamified VR experience group participants reported a clear sense of practice and control during specific operations: cutting the bitter melon (E4, E7, E11, E17), adding seasonings (E1, E5), and tearing out the pith and filling (E3, E12, E14, E16, E17).
Furthermore, the success of completion for the gamified VR experience group participants brought them a sense of achievement, such as \textit{``cutting bitter melon --- I place the knife and it cuts''} (E7) and \textit{``the knife drops, it separates, very responsive, leaps onto the plate --- there's a sense of accomplishment''} (E11). 
Therefore, successful gesture completion is called out and reinforced by gamified elements, such as step-completion audio and progress indicators, producing and sustaining micro-level mastery within bounded sub-tasks.

However, at the macro level, the gamified VR experience group participants still explicitly stated that they felt they were guided by the gamified VR experience. 
E8 articulated it: ``\textit{the order of operations is guided --- I just follow the game's program step by step.}''
Several others (E5, E6, E10, E13, E16) echoed this, specifically their reliance on the menu and bottle labels to proceed.
This contradiction between micro and macro levels shows a paradox where the micro-control makes the macro-constraint more visible: because participants felt in control of slicing, they noticed that they could not control the sequence in which slicing happened. 
E18 emphasized this by comparing to their real-world non-linear workflow: \textit{``I don't follow the recipe line-by-line. I mix some sauce, switch to cutting, return to the sauce; this matches my real-world workflow.''} 
This showed a design tension between participants enacting their own situated practice against the system mechanism, which purely observational interfaces cannot surface.

\subsubsection{Heritage Recognition Is Mode-Dependent: Narration Lands When Watching, Is Masked by Action}
\label{sec:heritage_recognition}

Both groups showed that the cultural narration works well in making them realize that stuffed bitter melon is a culinary ICH dish. 
For example, C1 stated directly: \textit{``no longer just `tasty'. I now grasp it embodies Hakka wisdom and inheritance, and want to explore further.''} 
C2 mentioned that \textit{``the `refined work on humble ingredients' wisdom''} and reported an attitude shift from pure consumption of the dish to respect. 
E7 stated that \textit{``through this experience, I learned this dish, which is an ICH from the Hakka region and the history about Hakka culture.''}

\begin{figure}[h]
  \centering
  \includegraphics[width=\linewidth]{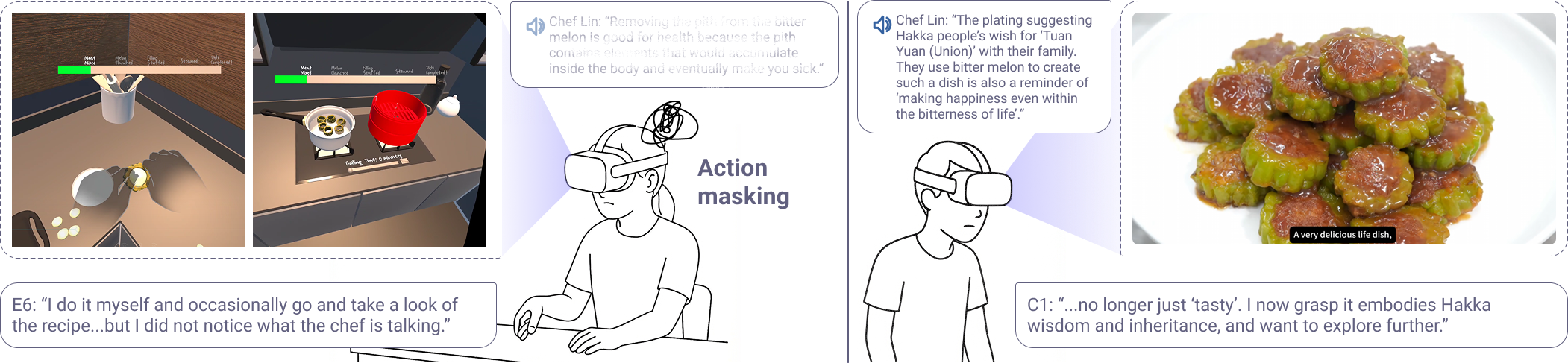}
  \caption{The Cultural Trade-off. A divergence between somatic attention and semantic uptake. The high cognitive load of enactment in the gamified VR experience condition caused ``action masking,'' where explicit cultural facts were missed in favor of somatic respect for the labor. In contrast, the Video condition facilitated narrative absorption, allowing participants to internalize the historical and symbolic context (e.g., Hakka spirit) through the auditory channel.}
  \label{fig:action}
\end{figure}

However, in the gamified VR experience group, the cultural narration was sometimes missed, echoing action masking (defined in Section~\ref{embodied_cognition})~\cite{tomporowski_cognitive-motor_2020, daniel_j_simons_gorillas_1999, easterbrook_effect_1959}: the motor demands of enactment suppress concurrent encoding of the narration~\cite{tomporowski_cognitive-motor_2020, daniel_j_simons_gorillas_1999, easterbrook_effect_1959} (see Figure~\ref{fig:action}). 
Nine of the twenty gamified VR participants felt that their understanding of the intangible cultural heritage/Hakka context was diluted due to their actions. 
E6 mentioned that \textit{``I do it myself and occasionally go and take a look at the recipe...but I did not notice what the chef was talking about.''}
E1 reported that ``\textit{I was analyzing the complexity of the process... decomposing the steps. The narration was interesting, but I felt there wasn't enough context because I was focused on the `doing'.}''
However, E1 also reported the cultural narrative that \textit{``the circular shape signifies family reunion; through the preparation culture I felt the Hakka persistence''}, showing that different actions might mask different levels of absorption.
This converges with the quantitative results: the gamified group showed no gain on Localization (Section~\ref{localization}) and was the only group to miss the purely narrated cultural-symbolism item (17/20 vs.\ 20/20; Section~\ref{quiz}) --- the item whose content was carried by narration alone. By contrast, the blanching rationale, engaged through enacted trial-and-error, showed the reverse numerical pattern (14/20 vs.\ 10/20).
This pattern is consistent with cognitive-load and split-attention accounts~\cite{Mayer_2022, Chandler_Sweller_1991}: when working memory is occupied by sensorimotor coordination, simultaneous narrative input is degraded in encoding.

\subsubsection{Authenticity Trade-offs and Complementarity}
\label{sec:authenticity}

Both media fell short of real cooking, but along different axes. 
First, the physical dynamic gap for VR still hinders the sense of immersion. Participants from both groups flagged that the kitchen setting and displayed materials were too ``idealized'' compared to their real-life experience. For the gamified VR experience group, as E11 reported, \textit{``[the bitter melon] split responsively and evenly, ... also makes me feel so unreal compared to my actual cooking skills.''}
For the Video group, six participants felt the ``\textit{glossy look of the bitter melon }(C16)'', ``\textit{perfect-look plating} (C19)'' and ``\textit{solid stuffing} (C5)'' look too perfect and feel like the video creator intentionally picked the most perfect ones, reducing perceived authenticity.

The missing sensory channels also reduce realism in the virtual environment.
C2 mentioned ``\textit{the video shows the filling pressed firmly, but I can't feel the stickiness. The steamer hisses, but I can't feel the heat.}'' 
Several participants (C1, C8) echoed the tactile-thermal absence. 
The experiment group reported the inverse trade-off: enactment gained, physical fidelity lost. 
E5 described grasp-and-weight failures (\textit{``objects disappear when touched or pick something up there's no gravity''}) and E6 reported the knife handle was hard to hold.
Across both conditions, the chemosensory layer, such as smell, taste, and thermal, was reported missing. 
E17 articulated it most fully: \textit{``the biggest gap is taste and smell. I can't smell the bitter melon or pork, can't taste the seasoning.'' }
C9 echoed the smell gap that \textit{``it would be even better if I could smell the aroma; right now it's all picture''}.

Participants in both groups independently mentioned a complementary mode merging the video-in-VR and VR-gamified-experience condition. 
Video viewers spontaneously argued for added enactment (C2, C3, C5, C7, C8, C9). 
Gamified VR participants wished to add watch-while-doing (E10: \textit{``I'd rather watch while doing, to improve completion''}). 
C20 specifically said that ``\textit{passive learning is for the sake of future active doing --- if doing teaches equally well, I prefer doing}. '' 
This shows that participants each preferred one mode but converged on a design that interleaves the two, pairing observation with operation.

\section{Discussion}\label{sec:Discussion}

\subsection{Embodied Enactment as Cultural Sense-Making}
Our quantitative results show that embodied VR engages more than passive observation: the gamified VR condition produced higher Interest/Enjoyment ($\delta=0.81$) and Sensory \& Imaginative Immersion ($\delta=0.62$), both robust to Holm correction, with a suggestive advantage in Positive Affect ($\delta=0.46$, $p_{\text{Holm}}=.063$), consistent with prior embodied VR/AR ICH work on dance~\cite{su_embodied_2024} and porcelain craft~\cite{tan_case_2020}.
However, prior embodied VR-ICH systems focus on externalizing performative heritage:
works such as \textit{Digital Diabolo}~\cite{Kong_2024} externalize a skill whose cultural value lies in movement, so fidelity is naturally operationalized as trajectory quality or creative expression, and ``success'' depends on the aesthetic continuum.  
We use \textit{Hakka Kitchen} as a case study for culinary ICH, showing that the movement is instrumental to an outcome (the dish), success is closer to binary (the stuffed bitter melon holds its filling or collapses), and cultural meaning is encoded in threshold-bound constraints such as slicing width, filling tightness, and blanching duration, rather than in expressive form. 
This enactment of constraints distinguishes it from the enactment of expression that \textit{Digital Diabolo} and kindred performative ICH systems~\cite{He_2023, nie_kites_2024} instantiate. 
This also aligns with the micro and macro paradox reported in Section~\ref{sec:agency}: participants gained fine micro-control over individual gestures while the system's linear gating and physics limits made the macro-constraint newly visible.
There is indeed a tension between gating and physics simulation, which is implementation-shaped and calibratable, but it is also intrinsic to the domain, because the constraints that generate it are properties of the craft. 
Performative ICH systems do not surface this tension because their domains contain no outcome that can fail. 
This is why the enactment-of-constraints framing is load-bearing rather than cosmetic, illustrating a design tension that is invisible and unavoidable in culinary heritage.

This tension also becomes pedagogically productive, echoing Kapur's productive failure~\cite{kapur_productive_2008}, which shows that learners who generate and explore suboptimal solutions before receiving instruction outperform those taught directly, because the failure activates and differentiates relevant prior knowledge ahead of consolidation. 
We extend this by examining it in a somatic condition where the hands redo the tasks, while Kapur's failure is conceptual. 
As the data show, the trade-off is affective rather than competence-based: the strongest and most robust cultural-awareness gain was on Transmission ($\delta=0.70$, surviving Holm correction), with Vitality showing a consistent but, after correction, suggestive pattern ($\delta=0.45$), while Perceived Competence and procedural knowledge did not differ.
Productive failure thus transfers to the heritage domain not as a route to skill but as a route to valuing, which is a reframing of the construct and a boundary on what our design can claim. 
Moreover, Kapur's instruction follows failure, whereas our cultural narration is concurrent with it, and this concurrency is precisely what allows the narration to be masked.
The contrast with Bjork's desirable difficulties~\cite{metcalfe_memory_1994} illustrates that our difficulty is not arbitrary friction introduced to slow encoding, but a culturally motivated constraint whose content is the lesson. 
Participants' reported ``respect'' for this dish is from experienced failure, since the video-in-VR group participants who only imagined the error did not develop the somatic-affective anchor that the gamified VR experience group participants did. 
That differentiation between anticipated and experienced constraint is the evidence that the mechanism is enactive, aligning with Reynolds' kinesthetic empathy~\cite{reynolds_kinesthetic_2012}, in which the user's physical struggle mirrors the artisan's labor and converts historical distance into affective proximity.

\subsection{Heritage Recognition is Mode-Dependent}
The same cultural narration reliably shifted attitudes in the video-in-VR condition but was 
frequently missed in the gamified VR experience condition. 
Nine of the gamified VR experience participants reported that their understanding of the ICH context was masked by their actions. 
This finding complicates the dominant assumption in the VR-ICH field that interactive embodied 
media straightforwardly improves cultural transmission~\cite{ji_constructing_2021, su_embodied_2024, tan_case_2020}.
The quantitative results corroborate this: the gamified VR condition raised the dimensions tied to valuing and continuity but produced no gain on Localization and numerically lower accuracy on the purely narrated cultural-symbolism item (85\% vs.\ 100\%; Section~\ref{quiz}) --- the one quiz item whose content was carried only by narration rather than by an enacted constraint. Conversely, the blanching rationale, which participants engaged through trial-and-error on the withheld-parameter timer, was the item on which the gamified group scored numerically highest relative to video (70\% vs.\ 50\%). Neither per-item difference reached significance, but the pattern dissociates in the predicted direction: enacted constraints were retained, while narrated-only meaning was masked.
Participants thus came to value the tradition and want to pass it on without necessarily encoding its situated, place-based detail, which is consistent with our popularization positioning and with ICH education goals that emphasize appreciation and continuity~\cite{heritage2003}. 
Even within China, culinary ICH is an intra-cultural transmission problem: our participants were subcultural novices to Hakka tradition, and came to recognize it as heritage only through the experience.
Similar to how social video platforms create heritage absorption through narrative~\cite{paquienseguy_douyin_2025, wang_critical_2024}, observation creates breadth of interest while embodied enactment anchors it in the body, but the two do not co-occur for free. 
This division of labor reframes prior VR-ICH evaluations, which report cultural-understanding gains from a single immersive condition; without a matched passive control, they cannot separate embodiment from repeated exposure to the content itself. 
\textit{FloraJing}~\cite{wang_facilitating_2025} reports that VR promotes progressive reflection and sustained cultural understanding, but evaluated only existing practitioners using VR, with no non-VR comparison, so its cultural gain is confounded by design. 
Our study then reveals that embodiment is not uniformly additive for cultural uptake. 
The two modes transmit different layers: video carries the declarative/narrative layer (framing, heritage significance) reliably, VR enactment carries the tacit/somatic layer (threshold parameters, felt constraints) reliably, and neither transmits both at once.

Failure of co-transmitting across modalities can be explained by cognitive-load and split-attention accounts~\cite{Chandler_Sweller_1991} concerning competition within a modality (e.g., two visual sources), and their canonical remedy, the modality principle~\cite{Mayer_2022}, offloads narration to the auditory channel. 
Our narration was already auditory, yet it was still masked during manual phases. This indicates the bottleneck in embodied enactment is not modality-specific but amodal: sustained sensorimotor coordination consumes the central working-memory and attentional control~\cite{wickens_multiple_2008} that narrative encoding also requires, regardless of input channel.
E1 illustrates this nuance: 
the same participant missed narration during manual action yet grasped the symbolic meaning of the circular form (family reunion),  showing that absorption is not all-or-nothing but varies with instantaneous load.
This points to a design opportunity: if cultural narration were delivered during low-load pauses, such as while steaming, rather than during high-load manual operations such as stuffing, heritage uptake during enactment might improve without sacrificing the somatic benefits of embodied practice.
Part of this mode-dependence is a fixable synchronization problem that better narration scheduling would mitigate. 
However, it cannot dissolve the effect for a structural reason.

The high-load manual phases are not incidental, but are the moments at which the culturally meaningful constraints are enacted, which means the periods of richest somatic learning are intrinsically the periods of poorest narrative uptake.
There is a tension between doing the constrained action and hearing why it matters: one that better timing can mitigate but not remove. 
This is also why the result is a property of the heritage domain rather than of our prototype: it should be most acute in threshold-bound, procedure-heavy crafts where manual coordination is dense, and least acute in expressive heritage like dance~\cite{su_embodied_2024} and kite-flying~\cite{nie_kites_2024}, where the movement itself carries the narrative, and there is no separate stream to mask.

\subsection{Design Implications}
Our study positions \textit{Hakka Kitchen} as a cultural awareness and popularization tool for process-based culinary ICH, designing meaningful enactment that helps players interpret why the practice matters while experiencing key constraints that constitute the heritage.

\subsubsection{Design the enactment of constraints, not just the procedure}

Experts define a practice as much by what not to do as by what to do. Treating expert-elicited negative constraints, such as the pith that must be fully scraped, as the primary drivers of gamified VR experience mechanics lets players learn by struggling against the same bottlenecks that carry the practice's cultural identity~\cite{metcalfe_memory_1994, lave_situated_1991}, so the player's struggle against the engine mirrors the artisan's against the material. This preserves the culturally salient friction that frictionless ``cooking games'' smooth away.

\subsubsection{Make feedback externalize the tacit cue, not just signal correctness}

Players anchor memory on action–feedback loops, repeatedly checking parameter ranges and outcome cues at key steps. Feedback should therefore surface the judgment being made, showing why a slice is too thick, not only that it is, through synchronized multimodal cues and a brief post-recovery rationale linking the constraint to its meaning. Because small mismatches between simulated and real cues reduced immersion (Section~\ref{sec:authenticity}), fidelity effort should concentrate on the specific cues that carry procedural judgment~\cite{persky_olfactory_2020, mazursky_thermalgrasp_2024}.

\subsubsection{The somatic bridge for cultural novices}

Even without full sensory fidelity, having the user's gesture spatially match the cultural action conveyed felt, first-person knowledge of the practice's constraints that observation did not.

The kinematic alignment of the hand created a virtual sense of competence beyond an effect of repetition: the video-in-VR group could pause and replay the identical content yet did not acquire this felt-parameter knowledge (Section~\ref{sec:knowing-about}), attributing the contribution to isomorphic enactment rather than mere exposure. 
Therefore, designers should also focus on the isomorphism of inputs, such as hand gesture tracking. The goal is to let the user perform the body of the practitioner, creating a somatic empathy that video cannot provide.

\subsubsection{Engagement strategies for ICH learning}
Serious games risk trading enjoyment for instructional intent~\cite{Dörner_Göbel_Effelsberg_Wiemeyer_2018, Laamarti_Eid_El_Saddik_2014}.
We addressed this by setting up playful items such as a progress indicator, an achievement badge, and a cultural story bonus, separating them from the ``ritual'' enactment of cooking, to support engagement without diluting the cultural and procedural authenticity of the experience.
However, post-play feedback revealed that a few participants still perceived the system as ``less like a game'' or ``not very enjoyable.'' 
To address this, future iterations could broaden the playful repertoire (e.g., cooperative or competitive cooking modes with AI-driven tasters) to deepen engagement without compromising heritage fidelity~\cite{ling_sketchar_2024, tang_breaking_2025, fu_being_2024, miller_eliciting_2025, he_i_2025}.

% refer to becoming my own audiences - listings 
\subsection{Limitations and Future Work}

\subsubsection{Attention Allocation: Action could mask cultural narration} 

Because cultural narration competes with manual load during high-load steps, step-boundary triggering did not fully prevent action masking.
Future designs should deliver narration in low-load windows or decouple it from action entirely, via dedicated reflection windows between phases. 
Notably, \textit{FloraJing}~\cite{wang_facilitating_2025} embedded cultural meaning spatially so that heritage cues are perceived peripherally without competing for the same attentional channel as manual action. 
This strategy avoids the split-attention problem our narration-during-action approach creates, and suggests that future culinary-ICH systems should explore environmental cultural framing alongside or instead of linear voiceover narration.
Our study also did not examine how players prioritize goals in a task-oriented environment.
Future work could also explore design solutions to rebalance attention between procedural tasks and cultural storytelling. 
One possible direction is a responsive, embodied Chef Lin who gestures, demonstrates, and reacts to the player's performance while telling the cultural stories in real time.
This embodied presence could capture players’ attention more effectively and reduce the tendency to deprioritize narrative in favor of task execution.

\subsubsection{Hint Usage Variability}

Optional hints create heterogeneous learning conditions where some learners may receive parameter ranges and rationales while others proceed via trial-and-error. 
This complicates attribution of knowledge outcomes.
Future studies should more systematically examine the role of hints by manipulating their availability or timing.
Logging and analyzing hint usage patterns could also clarify whether hints primarily reduce frustration, transmit deeper cultural knowledge, or simply act as optional aids without a strong impact on learning.

\subsubsection{Study Design}

Our evaluation was a single lab session with immediate, post-only measures, so it cannot speak to retention, real-kitchen transfer, or durability --- a constraint shared by most embodied VR-ICH studies~\cite{ji_constructing_2021, su_embodied_2024, tan_case_2020}. 
\textit{FloraJing}~\cite{wang_facilitating_2025} captured durability but lacked a matched passive control. 
Moreover, the short exposure duration and immediate testing window limited opportunities for consolidation, and the recall-oriented quiz items may not have fully captured the embodied and conceptual benefits of interactive learning.
Future work could combine our matched between-subjects design with a longitudinal deployment to test both our mode-dependence finding and the durability question.
This includes testing whether embodied VR practice supports better long-term procedural knowledge, deeper cultural appreciation, or real-world cooking uptake compared to passive in-VR video viewing. 
Complementary measures that probe conceptual understanding and transfer, beyond factual recall, would also provide a fuller picture of learning outcomes. 
Field studies in real-world kitchens could further reveal how lab-based outcomes translate into authentic settings and whether embodied interaction produces more sustainable impacts than observational formats.

\subsubsection{Control Design}
Most prior VR-ICH studies employ either no control condition or a non-VR baseline (desktop, verbal instruction, or conventional video)~\cite{ji_constructing_2021, 
su_embodied_2024, nie_kites_2024}, which conflates VR immersion with interaction modality. 
Our video-in-VR control holds the medium constant and varies only the interaction, isolating embodied enactment from VR immersion per se.
However, this design carries two limitations. First, although we selected footage that closely mimicked a chef’s viewpoint, mismatches in angle and framing remained compared to a true first-person perspective. Such discrepancies may have introduced perceptual bias, as participants in the control condition could interpret procedural steps differently than if they had viewed them directly from the chef’s eyes. Future control videos could be captured using head-mounted or stereoscopic equipment worn by the chef to ensure a more authentic alignment of perspective, thereby reducing perceptual mismatches and strengthening the validity of cross-condition comparisons.
Second, watching a seamless, error-free demonstration may inflate immediate procedural knowledge scores for the control group, as participants observe an ideal execution without the possibility of making mistakes. By contrast, interactive gameplay exposes participants to trial-and-error, which is theorized to foster more durable learning through embodied error-based practice~\cite{metcalfe2017}. This raises the possibility that our study may have underestimated the added value of interactivity when judged only by short-term outcomes. Future longitudinal studies should test whether the initial knowledge advantage of passive video viewing persists or diminishes over time compared to interactive practice, thereby clarifying whether VR’s trial-and-error learning supports more durable retention.

\subsubsection{Demographic Homogeneity}
\label{demographic_limitation}
Our participant sample consisted of individuals aged 18--37 years, all of Chinese background, with considerable variation in regional origin and time spent overseas.
Cross-cultural users (e.g., Western audiences) might struggle more with the basic affordances of the cuisine, potentially increasing cognitive load and action masking. 
Our participant sample was largely composed of individuals with higher-education backgrounds, which could be identified as a strong predictor of cultural participation and interest across diverse contexts~\cite{Heikkilä_2022}, 
potentially with elevated baseline interest or engagement, amplifying the gamified VR experience’s apparent effectiveness or masking differences visible in broader populations.
Future studies should recruit participants with more varied educational backgrounds and from different cultural contexts, which would help reveal whether engagement, knowledge, and awareness outcomes generalize across broader populations. It would also provide insight into how cultural familiarity or distance influences learning in VR-based culinary ICH experiences and how the somatic bridge holds across wider cultural distances.

\subsubsection{The ``Uncanny Valley'' of Multisensory VR Experience}

While \textit{Hakka Kitchen} preserves the choreography and visual logic of the cuisine, consumer VR cannot currently replicate the chemosensory cues that are central to professional cooking judgment. Systematic reviews confirm that few VR studies include any olfactory stimulus, and combined taste-smell-haptic conditions are rarer still~\cite{zholzhanova_virtual_2025}. 
Research-level prototypes demonstrate technical feasibility but not scalability ~\cite{mazursky_thermalgrasp_2024, liu_virchew_2025, persky_olfactory_2020}.
In a real kitchen, stuffing is guided by the tactile resistance of meat against the melon rind; doneness is judged by smell and steam behavior. In VR, players interact with weightless virtual objects and time-based sliders. Although we used visual and audio proxies, users inevitably miss the haptic and thermal feedback that tells a practitioner when a filling is too tight or too loose, so that they learn the motion but not the pressure~\cite{mazursky_thermalgrasp_2024}. 
This is the practical manifestation of the chemosensory representational gap that current consumer VR cannot approximate.

If consumer VR incorporated congruent olfactory cues, the threshold-bound property of culinary ICH would become directly testable through smell rather than a timer: practitioners traditionally judge steaming progress by aroma shift, not by clock.
The popularization-transfer boundary we accept would shift: with chemosensory feedback, VR could plausibly support not only cultural awareness but also elements of vocational judgment currently beyond its reach. 
However, the inverse-realism effect we observed (participants flagging ``too perfect'' visuals) suggests that poorly calibrated multisensory cues could backfire, in that an incongruent synthetic aroma might disrupt presence more than its absence~\cite{persky_olfactory_2020}.
This positions our work relative to non-VR culinary-ICH approaches that do not face the same sensory ceiling.  In-person co-design workshops with Hakka practitioners \cite{liu_salt_2025} made all chemosensory channels available, enabling participants to taste and judge the fidelity of GenAI-suggested recipe modifications; similar physical workshops \cite{chen_holocook_2025} allowed practitioners to handle real ingredients while 
imagining alternatives. These approaches trade scalability for sensory completeness. \textit{Hakka Kitchen} makes the inverse trade-off: scalable access at the cost of chemosensory fidelity, bounded to popularization rather than vocational transfer.

Future work should explore intermediate solutions: tracked physical proxies (knife handle, bowl, steamer lid) aligned with virtual objects to restore grasp affordances and tool semantics~\cite{10.1145/191666.191821}; photoreal 360° backgrounds to increase contextual plausibility and reduce the ``clean-room VR'' sensation~\cite{slater_place_2009, ritter_three-dimensional_2022}; and ambient scent diffusion (even non-precisely timed) to 
provide atmospheric olfactory context without requiring per-step synchronization.

\subsubsection{Scope and transferability}
Our study argues that the workflow of \textit{Hakka Kitchen} could be generalized under explicit boundary conditions.
On the design and method level, the following components are designed to be reusable:
(1) a practitioner-elicited procedural dictionary that identifies tacit checkpoints and failure modes;
(2) the enactment-of-constraints framework that ties those checkpoints to recoverable mechanics and feedback rules;
(3) the matched-media evaluation approach that isolates enactment from observation while keeping content aligned.
However, the following require redesign for each new recipe:
(1) the specific sensory markers and thresholds;
(2) tool semantics and their physics tuning;
(3) sociocultural scripts and meanings.

\section{Conclusion}\label{sec:Conclusion}

This work investigated how a gamified VR experience can support cultural awareness of process-based culinary ICH by making tacit constraints enactable. 
Across a matched between-subjects study, 
enactment increased interest, sensory and imaginative immersion, and the heritage-awareness dimensions tied to valuing and continuity, while our central finding qualifies this picture: 
heritage recognition is mode-dependent, with cultural narration reliably absorbed during observation but masked by manual load during enactment (action masking), so the two modes transmit complementary layers of heritage.
Qualitative findings revealed that recoverable error--repair loops and step-level constraint--action mappings served as somatic memory anchors, encoding the dish's cultural values into bodily experience.
This work contributes a transferable process for translating tacit culinary heritage into embodied mechanics, a practitioner-elicited procedural dictionary operationalized through an enactment-of-constraints framework, together with a matched-media evaluation approach that isolates enactment from observation.
Our findings suggest three design directions for future culinary-ICH technologies: design around culturally meaningful constraints rather than removing difficulty; make feedback externalize the tacit judgment rather than merely signal correctness; and schedule cultural framing so it is not masked by action.
Together, immersive interactive media offer a scalable path to strengthening public appreciation of, and thereby safeguarding, diverse ICH practices across domains.

% \begin{acks}
% thanks.
% \end{acks}

\bibliographystyle{ACM-Reference-Format}
\bibliography{references}

\clearpage
\onecolumn 
\appendix
\section{Appendix} \label{sec:Appendix}
\subsection{Study Participants' Demographics}
\label{Demogrpahic}

\begingroup
\footnotesize
\setlength{\tabcolsep}{3.5pt}
\renewcommand{\arraystretch}{1.05}
\begin{longtable}{@{}llclll cccc@{}}
\caption{Demographic table. Gaming = self-reported weekly gaming hours; Cooking = cooking sessions per week.}
\label{tab:demographics}\\
\toprule
Condition & ID & Age & Gender & Education & Cultural background &
\makecell{Gaming\\(h/wk)} & \makecell{Cooking\\($\times$/wk)} &
\makecell{Prior\\VR$^{\dagger}$} & \makecell{Motion\\sickness$^{\ddagger}$}\\
\midrule
\endfirsthead
 
\multicolumn{10}{c}{\tablename~\thetable~--~\textit{continued from previous page}}\\
\toprule
Condition & ID & Age & Gender & Education & Cultural background &
\makecell{Gaming\\(h/wk)} & \makecell{Cooking\\($\times$/wk)} &
\makecell{Prior\\VR$^{\dagger}$} & \makecell{Motion\\sickness$^{\ddagger}$}\\
\midrule
\endhead
 
\midrule
\multicolumn{10}{r}{\textit{Continued on next page}}\\
\endfoot
 
\bottomrule
\multicolumn{10}{@{}p{0.97\linewidth}@{}}{\rule{0pt}{2.2ex}$^{\dagger}$Higher values indicate more frequent prior VR use (self-report). $^{\ddagger}$Higher values indicate greater motion-sickness susceptibility (self-report).}\\
\endlastfoot

    Gamified VR & E1  & 25 & Male   & Master's    & Wuxi, Jiangsu          & $\geq$7 & 3 & 2 & 2 \\
    Gamified VR & E2  & 25 & Male   & Master's    & Wuxi, Jiangsu          & 0--3    & 4 & 1 & 1 \\
    Gamified VR & E3  & 28 & Female & Master's    & Zhengzhou, Henan       & 0       & 3 & 3 & 3 \\
    Gamified VR & E4  & 30 & Female & Master's    & Suzhou, Jiangsu        & 0       & 6 & 2 & 1 \\
    Gamified VR & E5  & 27 & Male   & Bachelor's  & Anhui                  & $\geq$7 & 1 & 3 & 2 \\
    Gamified VR & E6  & 28 & Female & Bachelor's  & Wuxi, Jiangsu          & 0       & 1 & 1 & 3 \\
    Gamified VR & E7  & 22 & Male   & Bachelor's  & Anhui                  & $\geq$7 & 5 & 1 & 2 \\
    Gamified VR & E8  & 37 & Female & Bachelor's  & Shanxi                 & 0--3    & 2 & 1 & 1 \\
    Gamified VR & E9  & 27 & Female & Bachelor's  & Xi'an, Shaanxi         & 0--3    & 2 & 1 & 2 \\
    Gamified VR & E10 & 24 & Female & Bachelor's  & Guangdong              & 0       & 3 & 2 & 2 \\ % CHANGED: background "Shenzhen, Guangdong" -> "Guangdong" (raw says Guangdong only; Shenzhen belongs to E11/E15/E16)
    Gamified VR & E11 & 24 & Female & Master's    & Shenzhen, Guangdong    & 0--3    & 7 & 4 & 1 \\ % CHANGED: education Bachelor's -> Master's (raw)
    Gamified VR & E12 & 24 & Female & Master's    & Tianjin                & 0--3    & 4 & 1 & 2 \\ % CHANGED: education Bachelor's -> Master's (raw)
    Gamified VR & E13 & 25 & Female & Bachelor's  & Guangzhou, Guangdong   & 3--7    & 3 & 1 & 7 \\ % CHANGED: motion 5 -> 7 (raw)
    Gamified VR & E14 & 24 & Female & Master's    & Chengdu, Sichuan       & 0--3    & 6 & 6 & 1 \\
    Gamified VR & E15 & 28 & Female & Master's    & Shenzhen, Guangdong    & 0--3    & 6 & 2 & 3 \\ % CHANGED: motion 7 -> 3 (raw)
    Gamified VR & E16 & 24 & Male   & Master's    & Shenzhen, Guangdong    & 0--3    & 3 & 2 & 2 \\
    Gamified VR & E17 & 27 & Female & Bachelor's  & Tianjin                & 0--3    & 2 & 4 & 6 \\ % CHANGED: education Master's -> Bachelor's (raw)
    Gamified VR & E18 & 23 & Female & Master's    & Henan                  & 3--7    & 1 & 3 & 2 \\ % CHANGED: age 27 -> 23 (raw); verify screening note re: partial Hakka ancestry
    Gamified VR & E19 & 30 & Male   & Master's    & Shanghai               & 0--3    & 4 & 3 & 1 \\
    Gamified VR & E20 & 25 & Male   & Master's    & Hunan                  & 0--3    & 2 & 5 & 1 \\
    \midrule
    VR video & C1  & 25 & Male   & Bachelor's  & Beijing                & $\geq$7 & 1 & 1 & 1 \\
    VR video & C2  & 21 & Male   & Bachelor's  & Tianshui, Gansu        & 3--7    & 1 & 2 & 2 \\
    VR video & C3  & 25 & Female & Master's    & Henan                  & $\geq$7 & 7 & 2 & 5 \\
    VR video & C4  & 23 & Female & Bachelor's  & Huangshan, Anhui       & 3--7    & 2 & 1 & 2 \\
    VR video & C5  & 24 & Female & Master's    & Qingdao, Shandong      & $\geq$7 & 3 & 2 & 3 \\
    VR video & C6  & 22 & Male   & Bachelor's  & Wuhan, Hubei           & $\geq$7 & 2 & 1 & 3 \\
    VR video & C7  & 23 & Female & Bachelor's  & Suzhou, Jiangsu        & 0--3    & 5 & 2 & 3 \\
    VR video & C8  & 25 & Male   & Master's    & Dalian, Liaoning       & $\geq$7 & 2 & 2 & 1 \\
    VR video & C9  & 21 & Female & Bachelor's  & Beijing                & 3--7    & 3 & 3 & 1 \\
    VR video & C10 & 21 & Female & Bachelor's  & Beijing                & 0       & 2 & 4 & 1 \\
    VR video & C11 & 20 & Female & Bachelor's  & Fuzhou, Fujian         & 0       & 1 & 3 & 3 \\ % CHANGED: education Master's -> Bachelor's (raw)
    VR video & C12 & 22 & Female & Bachelor's  & Nanjing, Jiangsu       & 0--3    & 2 & 3 & 1 \\ % CHANGED: entire row was a duplicate of C14; restored raw values (gaming 0-3, cooking 2, prior VR 3)
    VR video & C13 & 22 & Male   & Bachelor's  & Xi'an, Shaanxi         & 0--3    & 4 & 3 & 1 \\ % CHANGED: prior VR 2 -> 3 (raw)
    VR video & C14 & 22 & Female & Bachelor's  & Nanjing, Jiangsu       & 0       & 6 & 1 & 1 \\
    VR video & C15 & 22 & Male   & Bachelor's  & Zhejiang               & 3--7    & 5 & 6 & 7 \\ % CHANGED: education Master's -> Bachelor's (raw)
    VR video & C16 & 18 & Male   & High school & Shandong               & 0--3    & 4 & 5 & 6 \\
    VR video & C17 & 24 & Female & Bachelor's  & Hunan                  & 0--3    & 6 & 5 & 4 \\
    VR video & C18 & 24 & Male   & Master's    & Wuxi, Jiangsu          & 0--3    & 2 & 4 & 2 \\
    VR video & C19 & 25 & Female & Master's    & Wuxi, Jiangsu          & 0--3    & 5 & 2 & 4 \\
    VR video & C20 & 25 & Female & Master's    & Wuxi, Jiangsu          & 0--3    & 6 & 1 & 2 \\

\end{longtable}
\endgroup

\subsection{Codebook}
\label{Codeboook}
\begingroup
\footnotesize
\setlength{\tabcolsep}{4pt}
\begin{longtable}{@{}>{\raggedright\arraybackslash}p{0.16\textwidth} >{\raggedright\arraybackslash}p{0.24\textwidth} >{\raggedright\arraybackslash}p{0.50\textwidth}@{}}
\caption{Qualitative Analysis Codebook. Mapping of raw participant quotes to sub-codes and aggregated major themes, derived from the revised findings.}
\label{tab:codebook}\\
\toprule
\textbf{Theme} & \textbf{Subtheme} & \textbf{Example Quote} \\
\midrule
\endfirsthead
 
\multicolumn{3}{c}{\tablename~\thetable~--~\textit{continued from previous page}} \\
\toprule
\textbf{Theme} & \textbf{Subtheme} & \textbf{Example Quote} \\
\midrule
\endhead
 
\midrule \multicolumn{3}{r}{\textit{Continued on next page}} \\
\endfoot
 
\bottomrule
\endlastfoot
% =====================  T1  =====================
\multirow{8}{=}{\raggedright\textbf{Knowing-About vs.\ Knowing-How}\\[2pt]\emph{Enactment converts procedure into consequential, felt knowledge.}}
 & Procedural sequence transmits well in both conditions
 & \textit{``The order from the VR tutorial is clear. [You] just follow step by step.''} (C3); both groups produced high accuracy on the Procedural Knowledge Quiz (Fig.~\ref{ccr}).\\
\cmidrule(l){2-3}
 & Limit of watching: ``eyes learn, hands don't''
 & \textit{``Watching videos is like scrolling food bloggers --- the eyes learn but the hands don't.''} (C2) \\
\cmidrule(l){2-3}
 & Quantification gap: ``appropriate amount'' has no referent in video
 & \textit{``I had no idea how much.''} (C9); \textit{``How much in practice is anyone's guess.''} (C10) \\
\cmidrule(l){2-3}
 & Watching oversimplifies, even for experienced cooks
 & Watching makes cooking \emph{look} simple in a way that overpromises against real execution (C3, C20). \\
\cmidrule(l){2-3}
 & Tacit parameters become bodily-legible through enactment
 & \textit{``Through doing you get a direct impression of width, not just a flat `1--3 cm' on paper.''} (E1) \\
\cmidrule(l){2-3}
 & Embodied movement and spatial routine aid encoding
 & \textit{``Acting myself.''} (E4); \textit{``Laying out materials on the kitchen table and following the kitchen circulation route helps me memorize better.''} (E13) \\
\cmidrule(l){2-3}
 & Consequential, recoverable failure produces durable memory
 & \textit{``Self-operation makes memory stick because I fail and redo.''} (E17); error hints triggered by mistakes during doing reinforced filling-quantity and seasoning recall (E1). \\
\cmidrule(l){2-3}
 & Anticipated-but-unrealized error in video group
 & \textit{``I didn't make any errors, I was only watching, but I can imagine over-stuffing.''} (C1) \\
\midrule
% =====================  T2  =====================
\multirow{7}{=}{\raggedright\textbf{Agency and Pacing}\\[2pt]\emph{Embodied doing enabled micro-control but exposed macro-constraint.}}
 & Micro-agency: cutting the bitter melon
 & \textit{``Cutting bitter melon --- I place the knife and it cuts.''} (E7); \textit{``The knife drops, it separates, very responsive, leaps onto the plate --- there's a sense of accomplishment.''} (E11); also reported by E4, E17. \\
\cmidrule(l){2-3}
 & Micro-agency: adding seasonings
 & Sense of practice and control during the seasoning stage (E1, E5). \\
\cmidrule(l){2-3}
 & Micro-agency: scooping the pith and filling
 & Sense of practice and control during scooping/filling stages (E3, E12, E14, E16, E17). \\
\cmidrule(l){2-3}
 & Achievement loop reinforced by step-completion audio and progress indicators
 & Successful gesture completion produced micro-level mastery within bounded sub-tasks (E7, E11). \\
\cmidrule(l){2-3}
 & Macro-flow gated by the system's linear sequence
 & \textit{``The order of operations is guided --- I just follow the game's program step by step.''} (E8) \\
\cmidrule(l){2-3}
 & Reliance on menu and bottle labels to advance
 & Echoed reliance on the menu and bottle labels (E5, E6, E10, E13, E16). \\
\cmidrule(l){2-3}
 & Real-world workflow reasserts against linear gating
 & \textit{``I don't follow the recipe line-by-line. I mix some sauce, switch to cutting, return to the sauce; this matches my real-world workflow.''} (E18) \\
\midrule
% =====================  T3  =====================
\multirow{3}{=}{\raggedright\textbf{Heritage Recognition is Mode-Dependent}\\[2pt]\emph{Narration lands when watching, masked by action.}}
 & Realization of a Hakka Culinary ICH
 & \textit{``No longer just `tasty'. I now grasp it embodies Hakka wisdom and inheritance, and want to explore further.''} (C1); \textit{``Through this experience, I learned this dish, which is an ICH from the Hakka region and the history about Hakka culture.''} (E7) \\
\cmidrule(l){2-3}
 & Action Masking
 & \textit{``I was analyzing the complexity of the process... decomposing the steps. The narration was interesting, but I felt there wasn't enough context because I was focused on the `doing.'\,''} (E1) \\
\cmidrule(l){2-3}
 & Within-participant disconfirming: same E1 also grasped the cultural narrative
 & \textit{``The circular shape signifies family reunion; through the preparation culture I felt the Hakka persistence.''} (E1) \\
\midrule
% =====================  T4  =====================
\multirow{6}{=}{\raggedright\textbf{Authenticity Trade-offs and Complementarity}}
 & Idealized materials reduce perceived realism (Game)
 & \textit{``[The bitter melon] split responsively and evenly... also makes me feel so unreal compared to my actual cooking skills.''} (E11) \\
\cmidrule(l){2-3}
 & Idealized visuals reduce perceived realism (Video)
 & \textit{``Glossy look of the bitter melon''} (C16); \textit{``perfect-look plating''} (C19); \textit{``solid stuffing''} (C5). \\
\cmidrule(l){2-3}
 & Video preserves visual fidelity but loses touch and heat
 & \textit{``The video shows the filling pressed firmly, but I can't feel the stickiness. The steamer hisses, but I can't feel the heat.''} (C2); echoed by C1, C8. \\
\cmidrule(l){2-3}
 & Game gains enactment but loses physical fidelity
 & \textit{``Objects disappear when touched, or pick something up there's no gravity.''} (E5); the knife handle was hard to hold (E6). \\
\cmidrule(l){2-3}
 & Shared chemosensory ceiling across both conditions
 & \textit{``The biggest gap is taste and smell. I can't smell the bitter melon or pork, can't taste the seasoning.''} (E17); \textit{``It would be even better if I could smell the aroma; right now it's all picture.''} (C9) \\
\cmidrule(l){2-3}
 & Combining both modes
 & \textit{``Passive learning is for the sake of future active doing --- if doing teaches equally well, I prefer doing.''} (C20) \\
\end{longtable}
\endgroup

\end{document}